\documentclass[11pt,a4paper]{article}

\usepackage{jheppub}
\usepackage[T1]{fontenc}
\usepackage{fix-cm}
\usepackage{lmodern}
\usepackage{amsmath,amssymb,mathtools,mathrsfs,empheq}
\usepackage{physics}
\usepackage{bbm}
\usepackage{bm}
\usepackage{hhline}
\usepackage{caption}

\newcommand\nn{\nonumber}

\makeatletter
\newcommand*{\rom}[1]{\expandafter\@slowromancap\romannumeral #1@}
\makeatother

\usepackage{tikz}

\usepackage{xcolor}

\newcommand\fft[2]{\frac{#1}{#2}}

\usepackage{tensor}

\newcommand\mA{\mathcal{A}}

\newcommand\mY{\mathcal{Y}}
\newcommand\mS{\mathcal{S}}

\newcommand{\Z}{\mathbb{Z}}
\newcommand{\C}{\mathbb{C}}
\newcommand{\R}{\mathbb{R}}
\newcommand{\M}{\mathcal{M}}
\newcommand{\I}{\mathcal{I}}

\newcommand{\ZZ}{\mathcal{Z}}

\title{How to Build a Black Hole out of Instantons}
\author{Rishi Mouland}

\affiliation{DAMTP, Centre for Mathematical Sciences, \\
University of Cambridge, Wilberforce Road
Cambridge CB3 0WA, UK}

\emailAdd{r.mouland@damtp.cam.ac.uk}

\abstract{
An often fruitful route to study quantum gravity is the determination and study of quantum mechanical models---that is, models with finite degrees of freedom---that capture the dynamics of a black hole's microstates. An example of such a model is the superconformal quantum mechanics of Yang-Mills instantons, which has a proposed gravitational dual description as M-theory on a background of the form $X_7\times S^4$. This model arises in the strongly-coupled limit of the BFSS matrix model with additional fundamental hypermultiplets, offering a route towards useful numerical simulation. We construct a six-parameter black hole solution in this theory, which is generically non-supersymmetric and non-extremal, and is shown to arise in an ``ultra-spinning'' limit of the recently-found six-parameter AdS$_7$ solution. We compute its thermodynamic properties, and show that in the supersymmetric limit the entropy and on-shell action match precisely the expected results as computed from the superconformal index of the quantum mechanics, to leading order in the supergravity regime. The low-lying spectrum thus provides access to the dynamics of near-extremal black holes, whose spectra are expected to receive strong quantum corrections. 
}

\begin{document}
	
\maketitle 

\section{Introduction}
\label{sec: intro}

\subsection{Some background}

An essential feature any theory of quantum gravity is the identification of black holes as an appropriate ensemble of quantum states. Since the seminal work of Strominger and Vafa \cite{Strominger:1996sh}, the community has had great success in doing precisely that, for a broad gamut of supersymmetric black holes in various dimensions and backgrounds. More recently however, attention has turned to non-supersymmetric black holes, to which most of the usual tricks do not apply. 

AdS space presents a particularly concrete setting to seek a microscopic description of black holes, since one has a holographically dual field theory to work with. The task at hand then generically boils down to computing the density of states of a strongly-coupled field theory at large $N$, which is no mean feat. For BPS black holes, we have a trick up our sleeve: the corresponding degeneracies of BPS states can be effectively probed through supersymmetric indices, which can in many cases be computed exactly and their asymptotics studied. Through this approach, a slew of successful microstate countings have been achieved for BPS black holes in AdS in various dimensions \cite{Benini:2015eyy,Cabo-Bizet:2018ehj,Choi:2018hmj,Benini:2018ywd,Choi:2018fdc,Hristov:2018spe,Choi:2019dfu,Benini:2016hjo,Closset:2016arn,Hosseini:2017fjo,Benini:2017oxt,Hosseini:2016cyf,Gang:2019uay,Bobev:2019zmz,Hosseini:2017mds,Hosseini:2018dob,Nian:2019pxj,Choi:2019miv,Kantor:2019lfo,Nahmgoong:2019hko}\footnote{These citations are nowhere near exhaustive; see for instance the helpful review \cite{Zaffaroni:2019dhb} for further references.}. In contrast, a comparable microscopic derivation of the entropy of non-supersymmetric black holes requires access to dynamical (i.e. unprotected) observables in the strongly-coupled field theory, posing a much greater challenge. It is natural therefore to seek truncations or simplifying limits of holographic CFTs that still retain some information on non-BPS states of the full theory, while offering new opportunities for analytic and numerical study. A conceptually straightforward approach of this type is to truncate the CFT in radial quantisation to its zero modes on the sphere. When performed on a background suitable for a given BPS black hole, one arrives at a supersymmetric quantum mechanics whose Witten index by construction matches a suitable index in the initial theory. One thus expects the majority\footnote{That is, the number of BPS states in the quantum mechanical Hilbert space matches the number in the field theory to leading order in a large $N$ expansion. Discrepancies at lower orders can arise when one has to fix certain gauge fluxes on the sphere in performing the reduction \cite{Benini:2022bwa}.} of BPS microstates to survive as ground states in the quantum mechanics, while the low-lying spectrum should tell us something about near-BPS black holes. This approach has recently been applied to a particular topologically-twisted 3d $\mathcal{N}=2$ theory, suitable for the study of certain magnetically-charged black holes in AdS$_4$ \cite{Benini:2022bwa}.

A rather different, bulk approach to studying non-supersymmetric black holes has emerged in recent years, and is applicable for near-BPS and near-extremal black holes in various dimensions and backgrounds \cite{Sachdev:2015efa,Anninos:2017cnw,Nayak:2018qej,Hadar:2018izi,Larsen:2018cts,Sachdev:2019bjn,Castro:2019crn,Larsen:2020lhg,Charles:2019tiu,Hong:2019tsx,Moitra:2019bub,Castro:2018ffi,Moitra:2018jqs,Turiaci:2017zwd,Almheiri:2016fws,Sarosi:2017ykf,Iliesiu:2020qvm,Heydeman:2020hhw,Boruch:2022tno}. The key idea is that at leading order in a low-temperature expansion, the gravitational path integral is dominated by a theory of Jackiw-Teitelboim (JT) gravity in the AdS$_2$ throat of such black holes, supplemented by additional fields and couplings as dictated by the parent theory and black hole in question. This theory in turn generically exhibits a low-energy effective description in terms of a quantum mechanics with Schwarzian action, whose partition function can often be computed exactly. While subtle, this approach has offered novel insight into the microscopic entropy of near-BPS and near-extremal black holes, in doing so shedding new light on the long-standing black hole mass gap problem. In particular, one learns that the Bekenstein-Hawking entropy formula generically receives very strong quantum corrections at low temperatures.\\

There is a common theme that connects these two approaches: the realisation of a \textit{quantum mechanical} model that captures the dynamics of a black hole's microstates. In this paper, we study an example of a third approach to quantum gravity that shares this feature. From a bottom-up perspective, the idea is to propose a putative holographically dual gravitational description of a given non-relativistic conformal field theory (NRCFT) \cite{Nishida:2007pj,Goldberger:2008vg,Maldacena:2008wh,Son:2008ye,Balasubramanian:2008dm,Barbon:2008bg,Adams:2008wt,Herzog:2008wg}. Such field theories have as their spacetime symmetries the $z=2$ Schr\"odinger group Schr$(d)$ in $d$ spatial dimensions. Then, Schr$(d)$ can be realised as the maximal subgroup of the Lorentzian conformal group $SO(2,d+2)$ in $(d+1)$ spatial dimensions that commutes with a given null translation $P_+$, where the particle number operator $K$ is identified as $K=P_+$. Thus, at least in the simplest incarnation of the idea, one proposes a dual gravity theory living on a particular null compactification\footnote{The exact form of $X_7$ is given below; the straightforward generalisation to $X_{d+3}$ is given in \cite{Dorey:2023jfw}.} $X_{d+3}$ of AdS$_{d+3}$ of radius $R_\text{AdS}$. The isometries of $X_{d+3}$ form precisely Schr$(d)$. The conformal boundary of $X_{d+3}$ is the null-compactified pp-wave spacetime, as is suitable for an NRCFT in the oscillator frame \cite{Nishida:2007pj,Dorey:2023jfw}. It is worth noting that in general one should only expect a bulk semiclassical gravity description at large $K$ \cite{Maldacena:2008wh}, which is a regime that has been systematically studied for NRCFTs in some generality, most recently through the large charge programme \cite{Son:2005rv,Kravec:2018qnu,Kravec:2019djc,Orlando:2020idm,Pellizzani:2021hzx,Hellerman:2021qzz}.

Then, in a sector of fixed particle number $K$---or fixed momentum on the circle in the bulk---the NRCFT reduces to a conformal quantum mechanics. Thus, a black hole in the bulk with $K$ units of momentum must be recovered as a corresponding ensemble of states in this conformal quantum mechanics.

To test this idea, one would like to have a top-down construction of explicit dualities of this form. One route is to follow the original AdS/CFT mantra \cite{Maldacena:1997re}, except now we study the near horizon of a stack of $N$ spatially-compactified branes in string/M-theory in a limit in which the spatial compactification becomes null \cite{Maldacena:2008wh}. More generally, given any seed AdS/CFT dual pair, we can consider a particular limit on both sides \cite{Dorey:2023jfw}---a version of the limit of \cite{Seiberg:1997ad} relevant for CFTs---under which an AdS/NRCFT pair is in principle obtained. In the bulk, we end up with a gravitational theory formulated on $X_{d+3}\times \mathcal{N}$ for some compact space $\mathcal{N}$. Meanwhile, the dual NRCFT can be regarded as a \textit{definition} of the Discrete Lightcone Quantisation (DLCQ) of the initial theory on the branes \cite{Seiberg:1997ad}. However, such a procedure is in general ill-defined, due to the coupling of the theory to zero modes on the circle which become infinitely strongly-coupled in the null limit \cite{Hellerman:1997yu}\footnote{It may however be possible to avoid such pathologies for generic CFTs \cite{Fitzpatrick:2018ttk}, although we will not need these results here. }. 
 
There is one scenario however in which a sensible definition of the DLCQ is possible, and this is the case of $N$ M5-branes. Here, the high degree of supersymmetry along with atypical properties of the six-dimensional $(2,0)$ superconformal field theory living on the branes means that the DLCQ in a sector of $K$ units of momentum can be identified precisely as superconformal quantum mechanics (SCQM) on the moduli space $\M_{K,N}$ of $K$ Yang-Mills instantons in $SU(N)$ \cite{Aharony:1997th,Aharony:1997an}. The bulk theory on the other hand is M-theory on $X_7\times S^4$, which admits a semiclassical supergravity description in the regime $K\gg N^{7/3} \gg 1$ \cite{Maldacena:2008wh,Dorey:2022cfn}. In summary, we have the duality\vspace{0.4em}
\begin{align}\label{eq: duality}
  \begin{minipage}{60mm}\centering 
		Superconformal quantum mechanics of $K$ Yang-Mills instantons in $SU(N)$
	\end{minipage}
	\quad \longleftrightarrow \quad
	\begin{minipage}{60mm}\centering 
		M-theory on $X_7 \times S^4$, with $R_\text{AdS} = 2R_{S^4} = 2(\pi N)^{1/3}l_p$,\\ and $K$ units of momentum\\ on the circle in $X_7$
	\end{minipage}\\[-1em]\nn
\end{align}
It is important to emphasise that while this duality is motivated by considering the DLCQ of a conventional AdS$_7$/CFT$_6$ setup, it can be stated and studied in isolation:  as we shall see, there are things we can compute on the left-hand-side, and things we can compute on the right-hand-side, without recourse to the six-dimensional $(2,0)$ theory and DLCQ.

There are a number of ways to construct and study the quantum mechanics on the left-hand-side of (\ref{eq: duality}). While it can be thought of as simply a non-linear $\sigma$-model on $\M_{K,N}$, a perhaps more useful formulation is in terms of the D0-D4 system, with $K$ D0's and $N$ D4's. Viewing this system from the point of the view of the D0-branes, we precisely realise the SCQM as the strongly coupled limit of the Berkooz-Douglas matrix model \cite{Berkooz:1996is} in which dynamics is constrained to the Higgs branch $\M_{K,N}$. This matrix model is simply the $U(K)$ BFSS matrix model \cite{Banks:1996vh} coupled to $N$ fundamental hypermultiplets\footnote{See for instance Appendix A of \cite{Dorey:2022cfn} for more details}. Matrix models of this type have garnered renewed interest in recent years, as new technologies may begin to offer opportunities for their realistic simulation \cite{Gharibyan:2020bab,Rinaldi:2021jbg,Bauer:2022hpo,Maldacena:2023acv}. We can therefore view the bulk results of this paper---and in particular the presentation of a broad family of black hole solutions---as a detailed library of predictions that any such simulation must recover.

\subsection{The point of this paper}

The purpose of this paper is to construct black hole solutions on the right-hand-side of (\ref{eq: duality}). This amounts to constructing a black hole solution in seven-dimensional maximal gauged supergavity with $X_7$ asymptotics. The manifold $X_7$ is a particular null-compactification of AdS$_7$, defined as follows \cite{Dorey:2023jfw}. One can choose coordinates $(t,u,\xi, \rho_\alpha,\phi_\alpha)$, $\alpha=1,2$, on AdS$_7$ such that the metric takes the form
\begin{align}
  ds^2 = \frac{du^2}{g^2 u^2} + u^2 \left(-2 d\xi dt - \rho_\alpha\rho_\alpha  dt^2+ \sum_{\alpha=1}^2 \left(d\rho_\alpha^2 + \rho_\alpha^2 d\phi_\alpha^2\right) \right) - \frac{1}{g^2}dt^2
  \label{eq: X7 metric}
\end{align}
In particular, the conformal boundary ($u\to\infty$) realises the six-dimensional pp-wave geometry, where we've chosen two sets of polar coordinates $(\rho_\alpha,\phi_\alpha)$ in the transverse $\mathbb{R}^4 = \mathbb{R}^2\times \mathbb{R}^2$. The spacetime $X_7$ is defined by (\ref{eq: X7 metric}) subject to the null identification $\xi\sim \xi+2\pi$. In particular, this compactification breaks a quarter of the supersymmetry, so that $X_7\times S^4$ preserves 24 supercharges.

After constructing a broad family of the desired $X_7$ black hole solutions, we then verify that the on-shell action and Bekenstein-Hawking entropy of those that are supersymmetric is precisely recovered in an appropriate regime by the superconformal index of the quantum mechanics. In doing so, we provide a highly non-trivial check of the duality (\ref{eq: duality}).

\newpage 
\subsection{Summary of results}

In Section \ref{sec: initial solution} we detail our strategy. Rather than starting from square one, we start with the six-parameter AdS$_7$ solution found in \cite{Bobev:2023bxl}. This solution has three angular momenta, two electric charges, and an energy, and is generically non-extremal and non-supersymmetric. To arrive at our desired $X_7$ black hole, we follow a procedure outlined in \cite{Dorey:2022cfn}.

The basic idea is to consider a regime of the AdS$_7$ black hole solution in which one angular momentum is much larger than the others, $|\hat{J}_3/\hat{J}_\alpha|\gg 1$, $\alpha=1,2$. It has long been understood that we can in fact take the strict limit $|\hat{J}_3/\hat{J}_\alpha|\to \infty$, and still end up with a perfectly good supergravity solution. This is achieved through the introduction of a boosted coordinate system, which we can think of as rotating very quickly so as to `keep up' with the highly spinning black hole. This coordinate transformation looks locally like a Lorentz boost along a particular angle in $S^5$ with boost parameter $\beta=1-\eta^{-2}$, where $|\hat{J}_3/\hat{J}_\alpha|\sim \eta^2$. The reason why this highly-spinning regime is relevant here, is that the limit $\eta\to \infty$ is a Penrose limit in which the asymptotic geometry goes over from global AdS$_7$ coordinates to the $X_7$ metric (\ref{eq: X7 metric}), but without the identification $\xi\sim \xi+2\pi$.

To get this final ingredient, we need to refine the procedure slightly. We start once again with our AdS$_7$ black hole. Each radial slice of the AdS$_7$ black hole has $\mathbb{R}_t \times S^5$ topology. Then, the first step is to generalise this solution by orbifolding along the circle in the fibration $S^1\hookrightarrow S^5 \to \mathbb{CP}^2$, so that each slice is now $\mathbb{R}_t\times (S^5/\Z_n)$. We now take the combined limit $\eta,n\to\infty$, with $\eta^2 = n$. In doing so, the orbifold becomes simply $\xi\sim \xi+2\pi$ in the limit, and we obtain precisely $X_7$ asymptotics. In this limit, the angular momentum $\hat{J}_3$ and indeed also the energy $\hat{E}$ as measured in the initial frame scale like $\hat{E},\hat{J}_3\sim n$. However, the relevant thermodynamic properties for the limiting solution remain finite; these include the energy\footnote{We have set $g=1$ here for simplicity.} $E=\hat{E}+\hat{J}_3$ and momentum\footnote{In our conventions, we are taking $\hat{J}_3$ large and \textit{negative}. The converse case is trivially related.} $K = -\frac{1}{n}\hat{J}_3$, which correspond respectively to the scaling dimension and particle number of the NRCFT.

The end result is a broad family of asymptotically $X_7$ black holes. An illuminating summary of this construction is given below in Figure \ref{fig: ikea}.\vspace{0.7em}
\begin{center}
\begin{minipage}{\textwidth}\centering
\includegraphics[width=146mm]{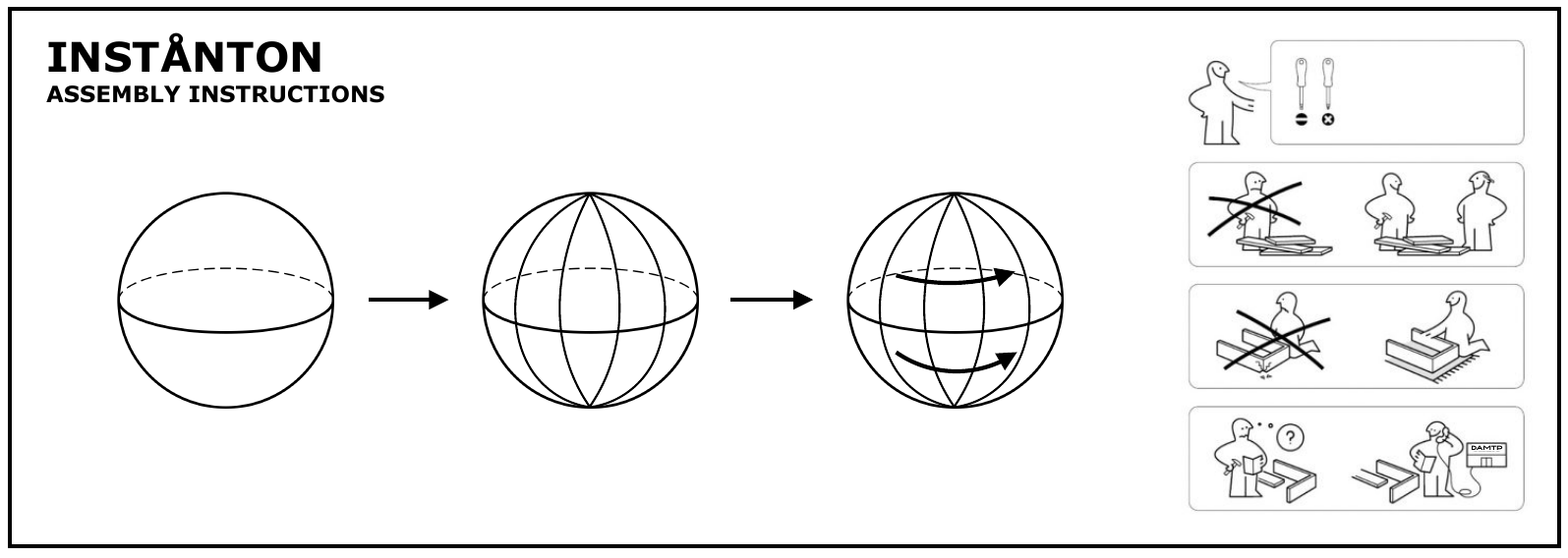}\vspace{2mm}
\begin{minipage}{140mm}\centering
\captionof{figure}{A schematic of the procedure to generate an asymptotically $X_{d+3}$ black hole \cite{Dorey:2022cfn}, here for $d=1$. Note, for $d$ odd, the orbifold necessarily has fixed points.}\label{fig: ikea}
\end{minipage}
\end{minipage}
\end{center}
\noindent This method can be viewed as a refinement of the limit first considered for precisely this purpose in \cite{Maldacena:2008wh}, where it was applied to AdS$_5$ black holes and a much simpler AdS$_7$ black hole. The essential idea there was to simply take $\eta\to\infty$, and at the end of the day impose the identification $\xi\sim \xi+2\pi$ by hand. Essentially the same limit was later considered in various dimensions and contexts \cite{Gnecchi:2013mja,Klemm:2014rda,Hennigar:2014cfa,Hennigar:2015cja,Wu:2021vch}, where it was dubbed the ``ultra-spinning'' limit\footnote{See also \cite{Emparan:2003sy,Caldarelli:2008pz,Caldarelli:2011idw} for related work. }. The end result is identical. The benefit of the formulation used here, however, is that for every finite $n$, we can unambiguously compute the thermodynamic properties of the black hole, which admit a smooth limit as $n\to\infty$. In this way, we resolve potential thermodynamic ambiguities encountered when computing thermodynamics in the ultra-spinning limit from scratch \cite{Hennigar:2014cfa,Hennigar:2015cja,Appels:2019vow}.

Note also that this same limit---the combined orbifold and Penrose limit---can be formulated on the field theory side. In particular, in \cite{Dorey:2023jfw}, the superconformal index of the SCQM was recovered in a limit of the index of the 6d $(2,0)$ theory on $S^1\times (S^5/\Z_n)$ \cite{Kim:2013nva}.\\

In Section \ref{sec: BH solution}, we present our solution in all its gory detail. This is a six-parameter black hole solution. Corresponding to the isometries of $X_7$, the black hole has an an energy $E$, two angular momenta $J_1,J_2$, and the all-important momentum $K$. It also has two electric charges $Q_1,Q_2$, corresponding to the isometries of the internal $S^4$. We present its thermodynamics, and discuss the limit in which it preserves some supersymmetry. Note that the equal electric charge case recovers the solution presented in \cite{Dorey:2022cfn}, which the present author produced there without derivation.\\

Finally, in Section \ref{sec: holography} we discuss holography. We first determine the range of parameter space in which the black hole solution can be trusted as a good M-theory background, which turns out to be $K \gg N^{7/3} \gg 1$. In this regime, the entropy (as well as the charges) of a typical black hole scales as
\begin{align}
  S\sim K^{1/2}N^{3/2}
  \label{eq: S scaling}
\end{align}
which must be reproduced by the density of states in the SCQM.

We are then able to precisely reproduce in the SCQM the entropy of supersymmetric black holes, for which the relevant task is to count BPS states. In detail, the leading order growth of the superconformal index of the SCQM in this regime is\footnote{The apparent discrepancy between the scaling of (\ref{eq: S scaling}) and (\ref{eq: I scaling}) is just an artefact of which ensemble we're in, since $\mu$ is a chemical potential for $K$ which scales as $K^{-1/2}N^{3/2}$ at the saddle point.} \cite{Dorey:2022cfn} 
\begin{align}
  -\log\mathcal{I} \sim   \frac{N^3}{24} \frac{\Delta_1^2 \Delta_2^2}{\mu \epsilon_1 \epsilon_2}
  \label{eq: I scaling}
\end{align}
in terms of various chemical potentials. We show that this result precisely matches the Euclidean on-shell action of the supersymetric black hole solution, as it must. We finally change ensemble, and show that the Bekenstein-Hawking entropy of the supersymmetric black holes,
\begin{align}
  S 	&= 2\pi\sqrt{\frac{3 Q_1 Q_2(Q_1+Q_2)-N^3 K(J_1+J_2)}{3(Q_1+Q_2)}} 
  \label{eq: intro S}
\end{align}
does indeed exactly match the growth of the superconformal index. In doing so, we derive a corresponding extremisation formula in the spirit of \cite{Hosseini:2017mds,Hosseini:2018dob}.\\

Some appendices provide details of a few calculations omitted from the main text.

\subsection*{A note on notation}\vspace{-0.8em}
\begin{align*}
  i			&=1,2,3		\quad &&\text{Three rotations in }S^5	\nn\\
  \alpha 	&=1,2		\quad &&\text{Two rotations in }\mathbb{R}^4 \text{ in pp-wave}			\nn\\
  I			&=1,2		\quad &&\text{Two rotations in internal }S^4, \text{ or equivalently Cartans of R-symmetry }SO(5)	
\end{align*}

\section{An AdS$_7$ black hole, an orbifold, and a limit}
\label{sec: initial solution}

Let us first set out our conventions for the supergravity theory we are looking for solutions in. We then briefly review the solution of \cite{Bobev:2023bxl}, and explain how it can be manipulated to provide the desired asymptotically-$X_7$ solution.

\subsection{The supergravity theory}

We work in a consistent truncation of 11-dimensional supergravity on $S^4$ \cite{Nastase:1999cb,Nastase:1999kf}. The solution of interest preserves a $U(1) \times U(1)$ subgroup of the $SO(5)$ symmetry of the gauged supergravity and therefore can be found in the $U(1) \times U(1)$ consistent further truncation of \cite{Liu:1999ai}. The bosonic fields in this consistent truncation are the metric, two Abelian gauge fields $A^I_{(1)}$, two real scalars $X_I$ and a single 3-form potential $A_{(3)}$ with a ``self-dual'' 4-form flux.

 The bosonic action of this $U(1) \times U(1)$ invariant consistent truncation in Lorentzian signature reads
\begin{align}\label{eq:S7d}
	S&=\fft{1}{16\pi G_N}\int\bigg[R \star 1 +2g^2\bigg(8X_1X_2+\fft{4(X_1+X_2)}{X_1^2X_2^2}-\fft{1}{X_1^4X_2^4}\bigg)\star 1 -\fft12\sum_{I=1}^2d\varphi_I\wedge \star d\varphi_I\nn\\
	&\kern6em~-\fft12\sum_{I=1}^2\fft{1}{X_I^2}F^I_{(2)}\wedge \star F^{I}_{(2)}-\fft{1}{2}X_1^2X_2^2F_{(4)}\wedge \star F_{(4)}\nn\\
	&\kern6em~+gF_{(4)}\wedge A_{(3)}+F^1_{(2)}\wedge F^2_{(2)}\wedge A_{(3)}\bigg]\,,
\end{align}
where we have defined
\begin{equation}
X_1 = e^{-\frac{1}{\sqrt{10}}\varphi_1-\frac{1}{\sqrt{2}}\varphi_2}\,, \quad X_2 = e^{-\frac{1}{\sqrt{10}}\varphi_1+\frac{1}{\sqrt{2}}\varphi_2}\,, \quad F_{(2)}^{I} = dA_{(1)}^{I}\,, \quad F_{(4)} = dA_{(3)}\,.
\end{equation}
It is straightforward to derive the bosonic equations of motion for this action. The Einstein equations read
\begin{equation}
\begin{split}
	0&=R_{\mu\nu}-\fft12g_{\mu\nu}\bigg[R+2g^2\bigg(8X_1X_2+\fft{4(X_1+X_2)}{X_1^2X_2^2}-\fft{1}{X_1^4X_2^4}\bigg)\bigg]\\
	&\quad-\sum_{I=1}^2\bigg[\fft12\partial_\mu\varphi_I\partial_\nu\varphi_I-\fft14g_{\mu\nu}\partial^\rho\varphi_I\partial_\rho\varphi_I\bigg]-\sum_{I=1}^2\fft{1}{X_I^2}\bigg[\fft12F^I_{\mu\rho}F^I_{\nu}{}^\rho-\fft18g_{\mu\nu}F^I_{\rho\sigma}F^{I\,\rho\sigma}\bigg]\\
	&\quad-X_1^2X_2^2\bigg[\fft{1}{12}F_{\mu\rho\sigma\lambda}F_{\nu}{}^{\rho\sigma\lambda}-\fft{1}{96}g_{\mu\nu}F_{\rho\sigma\lambda\delta}F^{\rho\sigma\lambda\delta}\bigg]\,.
\end{split}\label{eom:Einstein}
\end{equation}
The scalar equations of motion are given by
\begin{equation}
\begin{split}
	\Box\varphi_1&=\fft{1}{2\sqrt{10}}\sum_{I=1}^2\fft{1}{X_I^2}F^I_{\mu\nu}F^{I\,\mu\nu}-\fft{1}{12\sqrt{10}}X_1^2X_2^2F_{\mu\nu\rho\sigma}F^{\mu\nu\rho\sigma}\\
	&\quad+\fft{8g^2}{\sqrt{10}}\bigg(4X_1X_2-\fft{3(X_1+X_2)}{X_1^2X_2^2}+\fft{2}{X_1^4X_2^4}\bigg)\,,\\
	\Box\varphi_2&=\fft{1}{2\sqrt{2}}\bigg[\fft{1}{X_1^2}F^1_{\mu\nu}F^{1\,\mu\nu}-\fft{1}{X_2^2}F^2_{\mu\nu}F^{2\,\mu\nu}\bigg]+4\sqrt{2}g^2\fft{X_1-X_2}{X_1^2X_2^2}\,,
\end{split}\label{eom:scalar}
\end{equation}
where we have defined $\Box\equiv\nabla^\mu\nabla_\mu$. The vector and 3-form equations of motion take the form
\begin{subequations}
\begin{align}
	d(X_1^{-2} \star F^1_{(2)})&=F^2_{(2)}\wedge F_{(4)}\,,\label{eom:vector1}\\
	d(X_2^{-2} \star F^2_{(2)})&=F^1_{(2)}\wedge F_{(4)}\,,\label{eom:vector2}\\
	d(X_1^2X_2^2 \star F_{(4)})&=2gF_{(4)}+F^1_{(2)}\wedge F^2_{(2)}\,.\label{eom:3form}
\end{align}
\end{subequations}
In addition to the equations of motion, we must impose a ``self-duality'' equation on $A_{(3)}$, see \cite{Pilch:1984xy}, that reads
\begin{equation}
	X_1^2X_2^2 \star F_{(4)}=2gA_{(3)}+\fft12\left(A^1_{(1)}\wedge F^2_{(2)}+A^2_{(1)}\wedge F^1_{(2)}\right)-dA_{(2)}\,,\label{self-dual}
\end{equation}
written in terms of a 2-form potential $A_{(2)}$. Note that the existence of the 2-form potential $A_{(2)}$ satisfying the self-duality equation (\ref{self-dual}) can be shown locally by applying the Poincar\'{e} lemma to the 3-form equation of motion (\ref{eom:3form}), and therefore the self-duality equation does not impose any additional condition on a local solution to the equations of motion.

\subsection{The six-parameter AdS$_7$ black hole, and a tweak thereof}\label{subsec: orbifold}

A six-parameter asymptotically-AdS$_7$ solution to this supergravity theory was found in \cite{Bobev:2023bxl}, where the AdS radius is $R_\text{AdS} = g^{-1}$. Out of sympathy to the reader, we do not explicitly reproduce the solution here. We denote by $(\hat{t},\hat{r},\hat{y},\hat{z},\hat{\phi}_i)$ the coordinates used there. 

There is a simple generalisation of this solution that we can consider. The coordinates $(\hat{y},\hat{z},\hat{\phi}_1,\hat{\phi}_2,\hat{\phi}_3)$ span a topological $S^5$.
Then, for any integer $n$ and pair of integers $p_\alpha$, $\alpha=1,2$ that are each coprime to $n$, the black hole solution admits a free $\Z_n$ action, generated by
\begin{align}
  (\hat{t},\hat{r},\hat{y},\hat{z},\hat{\phi}_1,\hat{\phi}_2,\hat{\phi}_3) \longrightarrow  \left(\hat{t},\hat{r},\hat{y},\hat{z},\hat{\phi}_1+\frac{2\pi p_1}{n},\hat{\phi}_2+\frac{2\pi p_2}{n},\hat{\phi}_3 +\frac{2\pi}{n}\right)
  \label{eq: orbifold action}
\end{align}
For instance in the case $p_1=p_2=1$ we are simply performing discrete rotations along the fibre of the standard fibration $S^1 \hookrightarrow S^5 \to \C\mathbb{P}^2$.

Then by quotienting by this $\Z_n$ action we arrive at a new smooth black hole solution for each $(p_1,p_2)$, which we denote the $(n;p_\alpha)$ solution. The Lorentzian $(n;p_\alpha)$ solution has conformal boundary $\R_t \times \left(S^5/\Z_n[p_\alpha]\right)$, where $S^5/\Z_n[p_\alpha]$ denotes the smooth orbifold of $S^5$ given by $(\hat{y},\hat{z},\hat{\phi}_1,\hat{\phi}_2,\hat{\phi}_3)\sim (\hat{y},\hat{z},\hat{\phi}_1+\frac{2\pi p_1}{n},\hat{\phi}_2+\frac{2\pi p_2}{n},\hat{\phi}_3+\frac{2\pi}{n})$.

Computing the thermodynamic properties of the $(n;p_\alpha)$ solution is then straightforward, since it differs only globally from the original $n=1$ solution. The objects we'd like to compute are the energy $\hat{E}$, angular momenta $\hat{J}_i$, electric charges $\hat{Q}_I$, inverse Hawking temperature $\hat{\beta}=\hat{T}^{-1}$, angular velocities $\hat{\Omega}_i$, electrostatic potentials $\hat{\Phi}_I$, entropy $\hat{S}$, and on-shell action $\hat{I}$. Given some fixed $n$, all of these quantities are independent of the choice of $(p_1,p_2)$. So let $\hat{E}^{(n)}$, $\hat{J}_i^{(n)},\dots,\hat{I}^{(n)}$ denote the respective quantity for some $(n;p_\alpha)$ solution. We have then
\begin{align}
  \hat{E}^{(n)} &= \frac{1}{n} \hat{E}^{(1)},\quad & \hat{J}_i^{(n)} &= \frac{1}{n}\hat{J}_i^{(1)},\quad & \hat{Q}_I^{(n)} &= \frac{1}{n}\hat{Q}_I^{(1)}		\nn\\
  \hat{T}^{(n)} &= \hat{T}^{(1)} & \hat{\Omega}_i^{(n)} &= \hat{\Omega}_i^{(1)} &  \hat{\Phi}_I^{(n)} &= \hat{\Phi}_I^{(1)}
\end{align}
and
\begin{align}
  \hat{S}^{(n)} = \frac{1}{n} \hat{S}^{(1)},\qquad \hat{I}^{(n)} = \frac{1}{n}\hat{I}^{(1)}
\end{align}
The $n=1$ objects $\hat{E}^{(1)},\dots,\hat{I}^{(1)}$ are simply the thermodynamic properties of the solution of \cite{Bobev:2023bxl}. For brevity, we do not reproduce them here.

One can then verify that these thermodynamic quantities obey the first law of black hole thermodynamics,
\begin{align}
  d\hat{E}^{(n)} = \hat{T}^{(n)} d\hat{S}^{(n)} + \hat{\Omega}_i^{(n)} d\hat{J}_i^{(n)} +  \hat{\Phi}_I^{(n)} d\hat{Q}_I^{(n)}
\end{align}
where the variations are taken with respect to the parameters $(m,a_i,\delta_i)$. By construction, the on-shell action $\hat{I}^{(n)}$ obeys the quantum statistical relation \cite{Gibbons:1976ue}
\begin{align}
  \hat{I}^{(n)} = -\hat{S}^{(n)} + \hat{\beta}^{(n)}\left[\hat{E}^{(n)} - \hat{\Omega}_i^{(n)} \hat{J}_i^{(n)} -  \hat{\Phi}_I^{(n)} \hat{Q}_I ^{(n)}\right]
\end{align}

\subsection{Formulating the ultra-spinning limit}
\label{subsec: US limit}

Let us now demonstrate how the asymptotically-$X_7$ black hole we're after arises in the $n\to\infty$ limit of the $(n;p_\alpha)$ solution.

There exists an alternative set of coordinates $(t,u,\xi, \rho_\alpha,\phi_\alpha)$, $\alpha=1,2$ for the $(n;p_\alpha)$ black hole with the following properties. The $\phi_\alpha\sim \phi_\alpha + 2\pi$ are angles, while $\xi\sim \xi+2\pi n$ is also periodic. The pair $(\rho_1,\rho_2)$ provide coordinates on a quarter disc of radius $\sqrt{n}$, i.e. $\rho_\alpha\ge 0$ with $\rho_1^2 + \rho_2^2\le n$. Finally, the geometry is subject to the identification
\begin{align}
  (t,u,\xi,\rho_\alpha,\phi_\alpha) \sim  \left(t,u,\xi+2\pi,\rho_\alpha,\phi_\alpha+\frac{2\pi p_\alpha}{n} \right)
\end{align}
Working in these coordinates, as we approach the conformal boundary $u\to\infty$ we find the asymptotics
\begin{align}
  ds^2 = \frac{du^2}{g^2 u^2} + u^2 \bigg(& -2 d\xi dt - \rho_\alpha \rho_\alpha dt^2 + \sum_\alpha\left(d\rho_\alpha^2 + \rho_\alpha^2 d\phi_\alpha^2\right)		\nn\\
  & +\frac{1}{n} d\xi^2 + \frac{2}{n}\rho_\alpha\rho_\alpha d\xi dt + \frac{(\rho_\alpha d\rho_\alpha)^2}{\left(n-\rho_\alpha\rho_\alpha\right)} - \frac{1}{n^2}\rho_\alpha \rho_\alpha d\xi^2 \bigg) +\dots 
\end{align}
Appendix \ref{app: coordinates} provides details of how this coordinate system is defined with respect to the standard coordinates of the $(n;p_\alpha)$ solution. In particular, the limit $n\to\infty$ corresponds locally to a Penrose limit of the $\mathbb{R}\times S^5/\Z_n[p_\alpha]$ radial slices.

We then consider the solution found in the $n\to\infty$ limit. This solution has the desired $X_7$ asymptotics,
\begin{align}
  ds^2 = \frac{du^2}{g^2 u^2} + u^2  \left(-2 d\xi dt - \rho_\alpha \rho_\alpha dt^2 + \sum_\alpha\left(d\rho_\alpha^2 + \rho_\alpha^2 d\phi_\alpha^2\right)\right) + \dots 
  \label{eq: X7 asymptotics}
\end{align}
with $\phi_\alpha\sim \phi_\alpha +2\pi$ and $\xi \sim \xi+ 2\pi$ as required. However by explicit computation, one finds that if the parameters $(m,a_i,\delta_I)$ are held fixed as we take $n\to\infty$, the resulting solution is precisely the pure $X_7$ solution (\ref{eq: X7 metric}). This makes sense: as we take $n\to\infty$ we are moving to a highly rotating frame at infinity, while the angular momentum of the black hole is held fixed. In order to retain a black hole in the bulk, we need to take a compensating limit of large angular momentum in the direction in which our Penrose limit is taken. In particular, one needs to take
\begin{align}
  a_3 =  \frac{1}{g}\left(-1+\frac{1}{2\lambda n}\right)
  \label{eq: a3 behaviour}
\end{align}
for some free parameter $\lambda>0$. Then, holding fixed $(m,\lambda, a_\alpha,\delta_I)$ and taking the limit $n\to\infty$, we arrive at a six-parameter family of asymptotically-$X_7$ black hole solutions, which we detail in the following section.\\

Once uplifted to eleven dimensions, we thus find an asymptotically $X_7\times S^4$ solution. It would be interesting to independently study the stability properties of this $X_7\times S^4$ background, which we recall is a null compactification of AdS$_7\times S^4$ that preserves 24 supercharges. Another question is the background's causal structure; not least because it has a compact null circle. However as we shall see, for the black hole solutions we derive, this circle is blown up to be spacelike (and large in Planck units in a particular region) everywhere in the bulk, and only approaches vanishing invariant length as we approach the conformal boundary. One can thus imagine introducing a radial cutoff, with corresponding Fefferman-Graham expansion constrained such that the circle is everywhere strictly spacelike. Along these lines, it would be interesting to make contact with causality analyses of ten- and eleven-dimensional plane wave spacetimes (see e.g. \cite{Hubeny:2005qu,Marolf:2000cb}), although these spacetimes are rather different to (\ref{eq: X7 metric}) which is a compactification of a foliation of AdS$_7$ by six-dimensional plane wave spacetimes.

\section{The black hole solution}
\label{sec: BH solution}

Let us now present our six-parameter family of asymptotically-$X_7$ black hole solutions in full, and investigate its properties.

We first provide the metric and field content of the solution, and discuss global properties of the solution. We study the near-horizon geometry, and compute the thermodynamic properties of the solution. We finally study regions of parameter space for which the solution is supersymmetric.

\subsection{Details of the solution}

We present the solution in a coordinate system $(t,r,\xi,y,z,\phi_1,\phi_2)$, with $\xi\sim \xi+2\pi$ and $\phi_\alpha \sim \phi_\alpha + 2\pi$. The solution depends on six parameters $(m,\lambda,a_\alpha,\delta_I)$. The metric reads
\begin{equation}
	\begin{split}
		ds^2&= (H_1H_2)^{\fft15}\Bigg(-\frac{V}{g^2 \Xi_1^2 \Xi_2^2}dt^2 -\frac{2\lambda(1+g^2 r^2)(1-g^2 y^2)(1-g^2 z^2)}{g^2 \Xi_1 \Xi_2} dt \,d\xi\nn\\
		& \hspace{25mm}+\fft{(r^2+y^2)(r^2+z^2)}{U}dr^2\\
		&  \hspace{25mm}+\fft{g^2 y^2(r^2+y^2)(y^2-z^2)}{(1-g^2y^2)^2(a_1^2-y^2)(a_2^2-y^2)}dy^2 \\
		&  \hspace{25mm}+\fft{g^2z^2(r^2+z^2)(z^2-y^2)}{(1-g^2z^2)^2(a_1^2-z^2)(a_2^2-z^2)}dz^2 \\
		&  \hspace{25mm}+\fft{g^2(a_1^2+r^2)(a_1^2-y^2)(a_1^2-z^2)}{\Xi_1^2(a_2^2-a_1^2)}d\phi_1^2+\fft{g^2(a_2^2+r^2)(a_2^2-y^2)(a_2^2-z^2)}{\Xi_2^2(a_1^2-a_2^2)}d\phi_2^2  \\
		&\hspace{25mm}+\fft{1-\fft{1}{H_1}}{1-(s_2/s_1)^2}K_1^2+\fft{1-\fft{1}{H_2}}{1-(s_1/s_2)^2}K_2^2 \Bigg) \,, 
	\end{split}\label{ansatz:metric}
\end{equation}
where
{\allowdisplaybreaks
\begin{subequations}
\begin{align}
		s_I&=\sinh\delta_I\,, \quad c_I=\cosh\delta_I\,, \\
		\Xi_\alpha &= 1-a_\alpha^2g^2\,, \\
		H_I(r,y,z)&=1+\fft{2ms_I^2}{(r^2+y^2)(r^2+z^2)}\,,\\
		U(r)&=\fft{(1+g^2r^2)^2(r^2+a_1^2)(r^2+a_2^2)}{g^2r^2}-2m+m(s_1^2+s_2^2)(1+g^2(a_1^2 + a_2^2) + 2g^2r^2) \nn\\
		&\quad +\fft{4m^2g^2s_1^2s_2^2}{r^2} +\fft{2m(s_1^2+s_2^2)a_1a_2}{r^2}\nonumber \\
		&\quad+ \frac{m(c_1-c_2)^2}{2r^2}\left(r^2+(a_1-a_2)^2\left(4-4a_1 a_2 g^2+4 g^2 r^2 - g^4(a_1+a_2)^2 r^2\right)\right)\,,\\
		V(r,y,z) &= 1-2g^2(a_1^2 + a_2^2) + g^4\left(r^2 y^2 + r^2 z^2 - y^2 z^2 + (a_1^2 + a_2^2 )(y^2 + z^2 - r^2)+3a_1^2 a_2^2\right)		\nn\\
		&\quad -2g^6 \left(r^2 y^2 z^2 + a_1^2 a_2^2 (y^2 + z^2 - r^2)\right) \nn\\
		&\quad+g^8\left((a_1^2 + a_2^2)r^2 y^2 z^2+ a_1^2 a_2^2(y^2 z^2 - y^2 r^2 - z^2 r^2)\right)		\\
		K_1&=\fft{c_1+c_2}{2s_1}\mathcal{A}[y^2,z^2,0]+\fft{c_1-c_2}{2s_1}\mY\,,\\
		K_2&=\fft{c_1+c_2}{2s_2}\mathcal{A}[y^2,z^2,0]+\fft{c_2-c_1}{2s_2}\mY\,,\\
		\mY&=-g^2(a_1-a_2)^2 \mA[y^2,z^2,0] +2(1-a_1 a_2 g^2) \mA[y^2,z^2,g^{-2}] \nn\\
		&\quad+  \frac{(a_1^2 - y^2)(a_1^2 - z^2)(2-2a_1 g^2 (a_1 + a_2)+a_1^2 g^4(a_1^2 + a_2^2))}{a_1 \Xi_1^2( a_1^2 - a_2^2)} d\phi_1		\nn\\
		&\quad+  \frac{(a_2^2 - y^2)(a_2^2 - z^2)(2-2a_2 g^2 (a_1 + a_2)+a_2^2 g^4(a_1^2 + a_2^2))}{a_2 \Xi_2^2( a_2^2 - a_1^2)} d\phi_2
	\label{ansatz:functions}
\end{align}
\end{subequations}
}
and for any $v_1,v_2,v_3$ we define the one-form
\begin{align}
  \mA[v_1,v_2,v_3] &=  \frac{1}{2g\Xi_1^2 \Xi_2^2}\bigg( 	(1-3 a_1^2 g^2 - 3 a_2^2 g^2 + 5 a_1^2a_2^2 g^4) \nn\\
  &\hspace{22mm} + g^2 (1 + a_1^2 g^2 + a_2^2 g^2 - 3 a_1^2 a_2^2 g^4)(v_1 + v_2 + v_3)				\nn\\
  & \hspace{22mm}+ g^4 (-3 + a_1^2 g^2 + a_2^2 g^2 + a_1^2 a_2^2 g^4)(v_1 v_2 + v_2 v_3 + v_3 v_1)			\nn\\
    & \hspace{22mm} +g^6 (5 - 3 a_1^2 g^2 - 3 a_2^2 g^2 + a_1^2 a_2^2 g^4)v_1 v_2 v_3 \bigg) dt	\nn\\
  &\quad + \frac{\lambda(1-g^2 v_1)(1-g^2 v_2)(1-g^2 v_3)}{g\Xi_1 \Xi_2}\,d\xi			\nn\\[0.5em]
  &\quad + \frac{g^2(a_1^2-v_1)(a_1^2-v_2)(a_1^2-v_3)}{a_1\Xi_1^2(a_1^2 - a_2^2)} d\phi_1 + \frac{g^2(a_2^2-v_1)(a_2^2-v_2)(a_2^2-v_3)}{a_2\Xi_2^2(a_2^2 - a_1^2)} d\phi_2
  \label{eq: B def}
\end{align}
The two Abelian gauge fields are given in terms of $K_{1,2}$ as
\begin{equation}\label{eq:A1Idef}
	A^I_{(1)}=\bigg(1-\fft{1}{H_I}\bigg)K_I + \alpha_I dt\,.
\end{equation}
for any constants $\alpha_I$, since the final term is pure gauge. The importance of including such a term will become clear shortly.

The 3-form is given by
\begin{equation}
	\begin{split}
		A_{(3)}&-2ms_1s_2g^3a_1a_2\bigg[\mA[y^2,z^2,0]-\mA[y^2,z^2,g^{-2}]\bigg]\\
		&\hspace{3mm}\wedge\bigg[\fft{dz\wedge\left(\mA[y^2,0,0]-\mA[y^2,0,g^{-2}]\right)}{(r^2+y^2)z}+\fft{dy\wedge\left(\mA[z^2,0,0]-\mA[z^2,0,g^{-2}]\right)}{(r^2+z^2)y}\bigg]\\
		&\hspace{3mm}+2ms_1s_2g^3\mA[y^2,z^2,0]\\
		&\hspace{3mm}\wedge\left[\fft{z \, dz\wedge\left(\mA[y^2,0,0]-\mA[y^2,0,g^{-2}]\right)}{(r^2+y^2)}+\fft{y\, dy\wedge\left(\mA[z^2,0,0]-\mA[z^2,0,g^{-2}]\right)}{(r^2+z^2)}\right]\,,\label{ansatz:3form}
	\end{split}
\end{equation}
The two scalars are given by
\begin{equation}
	X_I=\fft{(H_1H_2)^\fft25}{H_I}\,.\label{ansatz:scalar}
\end{equation}
Finally, the self-duality constraint (\ref{self-dual}) is satisfied by the 2-form
\begin{align}
  A_{(2)} &= \left(\frac{1}{H_1}+\frac{1}{H_2}\right)\frac{ms_1 s_2}{(u^2 + y^2)(u^2 + z^2)} \mA[y^2, z^2,0]  \nn\\
  &\qquad  \wedge \Bigg((1-a_1 a_2 g^2)\mA[y^2,z^2,g^{-2}]	+ \frac{(a_1^2 - y^2)(a_1^2-z^2)}{a_1(a_1^2-a_2^2)}d\phi_1+ \frac{(a_2^2 - y^2)(a_2^2-z^2)}{a_2(a_2^2-a_1^2)}d\phi_2\Bigg)			\nn\\
  &\qquad - \frac{1}{2}dt\wedge \left(\alpha_1 A^2_{(1)} +\alpha_2 A^1_{(1)}\right)
  \label{eq: 2-form solution}
\end{align} 
where recall the $\alpha_I$ are as-of-yet undetermined constants appearing in (\ref{eq:A1Idef}).

This solution then admits an alternative set of coordinates $(t,u,\xi,\rho_\alpha,\phi_\alpha)$ such that as we approach the conformal boundary ($u\to\infty$) we realise the desired $X_7$ asymptotic geometry (\ref{eq: X7 asymptotics}). The relation between these coordinates and the coordinates $(t,r,\xi,y,z,\phi_1,\phi_2)$ in the solution as presented above is given in (\ref{eq: X7 coord change}).\\

We should be a little more precise about the range of coordinates. The rotation parameters $a_1,a_2$ lie in the range $-1\le a_\alpha g \le 1 $. Assuming without loss of generality that $a_1^2 < a_2^2$, the coordinates $y,z$ run over non-negative values such that
\begin{align}
  a_1^2 \le z^2 \le a_2^2 \le y^2 \le g^{-2}
\end{align}
Meanwhile the $\phi_\alpha\sim \phi_\alpha + 2\pi$ are just angles on a $T^3$, and $\xi\sim \xi+2\pi$ is also periodic. The radial coordinate runs over $r_+ \le r < \infty$, where $r\to \infty$ is the asymptotically $X_7$ boundary region, while $r=r_+$ is the outer horizon, defined implicitly as the largest positive root $U(r_+)=0$ of $U(r)$. In many of the expressions below, we keep $r_+$ as an implicit function of the parameters $\{m,\lambda,a_\alpha, \delta_I\}$. In Lorentzian signature the time coordinate $t$ is non-compact, but we will often work in Euclidean time $\tau$, defined by
\begin{align}
  t=-i \tau
\end{align}
We should finally comment on the topology of the solution. A slice of constant time and radial distance, such as the outer horizon, has the topology $S^1 \times \R^4$, and so is in particular non-compact. Nonetheless, the horizon has finite area. Indeed, this is a standard feature of an ultra-spinning black hole, as first noted in \cite{Klemm:2014rda}. In general, an ultra-spinning black hole in $D$ dimensions will have a horizon topology $S^{D-2}- S^{D-4}$, where in this case $S^1\times \R^4 = S^5 - S^3$.

Let us provide a schematic explanation of how this topology arises in the limit formulated in Section \ref{sec: initial solution}. At each finite $n$, each constant time and radius slice has topology $S^5/\Z_n$, where $\Z_n$ acts along the fibre $S^1 \hookrightarrow S^5 \to \mathbb{CP}^2$. We can then in turn view $\mathbb{CP}^2$ as a $T^2$ fibration over a triangle, where the torus degenerates to a circle on each side, and to a point at each vertex; see Figure \ref{fig: toric}. It follows that in the neighbourhood of each corner, $S^5/\Z_n$ has the topology $S^1 \times \R^4$. Then, in the strict $n\to\infty$ limit, we in fact change the topology of the spacetime. The Penrose limit zooms into an open neighbourhood of the third corner of the triangle, while the orbifold ensures that the Hopf fibre does not decompactify. We are essentially cutting the geometry along the dotted line; the excised two-node toric diagram is precisely $\mathbb{CP}^1=S^2$, and so once we include the additional non-trivial circle fibre over $\mathbb{CP}^2$, we are ultimately excising the lens space $S^3/\Z_n$ from $S^5/\Z_n$. As a result, the limiting spacetime has fixed time and radius slices with topology $(S^5/\Z_n)- (S^3/\Z_n) \cong S^5- S^3 \cong S^1 \times \R^4$.
\begin{center}
\begin{minipage}{\textwidth}
\centering
\includegraphics[width=80mm]{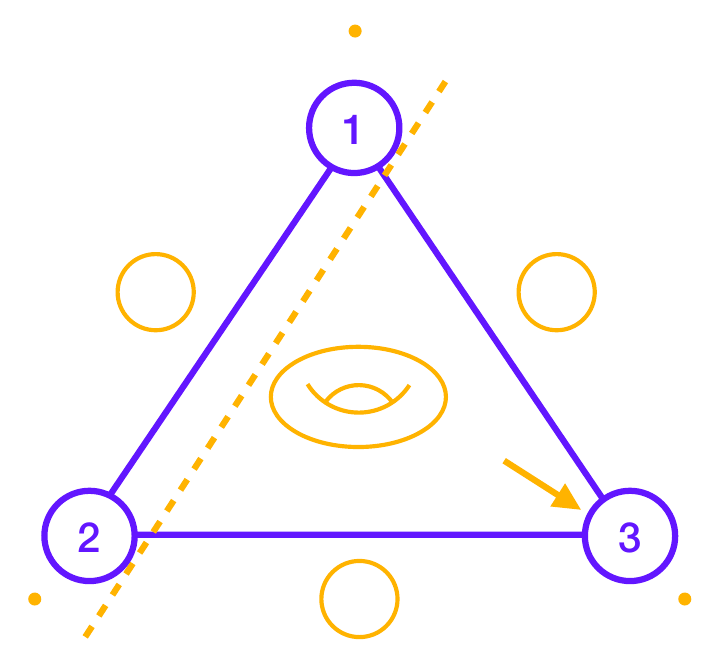}
\captionof{figure}{The toric diagram of $\mathbb{CP}^2$. In the strict $n\to\infty$ limit, we excise two nodes which form $\mathbb{CP}^1=S^2$. }\label{fig: toric}
\end{minipage}
\end{center}

\subsection{Near-horizon geometry}

Next, let us study the near-horizon geometry. We define a new coordinate $\sigma$ by
\begin{align}
  r=r_+ + \sigma^2
\end{align}
Then, in the regime $\sigma^2\ll r_+$, the metric takes the form
\begin{equation}
\begin{split}
	ds^2\big|_{\sigma^2\ll r_+}&=4 U'(r_+)^{-1}(H_1H_2)^{\fft15}(r_+^2+y^2)(r_+^2+z^2)\bigg[d\sigma^2 + \left(\frac{2\pi}{\beta}\right)^2\sigma^2d\tau^2\bigg]\nn\\
	&\quad+(H_1H_2)^{\fft15}\fft{(r_+^2+y^2)(y^2-z^2)y^2}{(1-g^2y^2)(a_1^2-y^2)(a_2^2-y^2)(a_3^2-y^2)}dy^2\\
	&\quad+(H_1H_2)^{\fft15}\fft{(r_+^2+z^2)(z^2-y^2)z^2}{(1-g^2z^2)(a_1^2-z^2)(a_2^2-z^2)(a_3^2-z^2)}dz^2\\
	&\quad+g_{\alpha\beta}(r_+,y,z)(d\phi_\alpha-i\Omega_\beta d\tau)(d\phi_\beta-i\Omega_\beta d\tau)\\
	&\quad+ g_\alpha(r_+,y,z)(d\phi_\alpha-i\Omega_\beta d\tau)(d\xi-i\Upsilon d\tau) \\
	&\quad+ g(r_+,y,z)(d\xi-i\Upsilon d\tau)^2\,,
\end{split}\label{metric:nH}
\end{equation}
which we have written in terms of the Euclidean time $\tau=it$. Here, $g_{\alpha\beta},g_\alpha$ and $g$ are complicated functions we will not need to explicit forms of, and the $H_I$ are evaluated at $H_I = H_I(r_+,y,z)$. The $\{\beta,\Upsilon,\Omega_\alpha\}$ are constants given below.

The regularity of the near horizon geometry at $\sigma=0$ determines the periodic identification of coordinates along the thermal circle as
\begin{align}
  (\tau,\xi,\phi_\alpha) \sim (\tau+\beta,\xi+i\beta\Upsilon,\phi_\alpha+i\beta \Omega_\alpha)
\end{align}
We can then read off some thermodynamic quantities. First, we identify the Hawking temperature as
\begin{align}
  T = \beta^{-1} = \frac{r_+^2 U'(r_+)}{4\pi g \sqrt{\mathcal{S}(r_+)}}
\end{align}
where the function $\mathcal{S}$ is defined as
\begin{align}
  g^4\mathcal{S}(r)&= \prod_{I=1}^2\Big((r^2+a_1^2)(r^2+a_2^2)(1+g^2r^2)+2mg^2s_I^2(r^2+a_1a_2)\Big)\\
	&\quad+2mg^2(c_1-c_2)^2(a_1-a_2)^2(1-a_1a_2g^2)(r^2+a_1^2)(r^2+a_2^2)(1+g^2 r^2)
	\label{eq: S def}
\end{align}
This same result can also be obtained by computing the surface gravity at the outer horizon; we omit the details of this calculation. 

Next, we have the angular velocities $(\Upsilon,\Omega_\alpha)$, which are equivalently defined as constants such that the Killing vector
\begin{align}
  l=\partial_t - \Upsilon \partial_\xi - \Omega_\alpha \partial_{\phi_\alpha}
\end{align}
is null at the outer horizon, i.e. $l^\mu l_\mu|_{r=r_+} =0$. We find
\begin{align}
  \Upsilon=\fft{1}{g^4 \lambda \mS(r_+)}\Bigg[&\frac{1}{2}(r_+^2+a_1^2)^2(r_+^2 + a_2^2)^2(1-g^4 r_+^4)\nn\\
  &-g^2 m(r_+^2 + a_1^2)(r_+^2 + a_2^2)(g^2 r_+^4 - a_1 a_2)(s_1^2 + s_2^2)- 2g^4 m^2(r_+^4 - a_1^2 a_2^2)s_1^2 s_2^2\nn\\
  &+m g^2 (c_1-c_2)^2(a_1-a_2)^2(1-a_1 a_2 g^2)(a_1^2 + r_+^2)(a_2^2 + r_+^2)\Bigg]
\end{align}
and
\begin{align}
	\Omega_1&=-\fft{1}{g^5\mS(r_+)}\Bigg[\frac{1}{2}\bigg((1+g^2r_+^2)(r_+^2 + a_1^2)(r_+^2 + a_2^2)+2mg^2 s_1^2(r_+^2+a_1a_2)\bigg)\nn\\
	&\hspace{28mm} \times\bigg(a_1(1+g^2r_+^2)^2(r_+^2+a_2^2)+2mg^2s_2^2(a_1g^2r_+^2+a_2)\bigg) \notag\\
	&\hspace{20mm}+\frac{1}{2}\bigg((1+g^2r_+^2)(r_+^2 + a_1^2)(r_+^2 + a_2^2)+2mg^2s_2^2(r_+^2+a_1a_2)\bigg)\nn\\
	&\hspace{28mm}\times \bigg(a_1(1+g^2r_+^2)^2(r_+^2+a_2^2)+2mg^2s_1^2(a_1g^2r_+^2+a_2)\bigg)\notag\\
	&\hspace{20mm}-mg^2(c_1-c_2)^2(a_1-a_2)^2(r_+^2+a_2^2)(1+g^2r_+^2)\nn\\
	&\hspace{28mm}\times \bigg(2(gr_+^2-a_1)(1-a_1a_2g^2)-g(1+a_1g)^2(2-a_1g-a_2g)r_+^2\bigg)\Bigg]\,.\label{eq:Omega}
\end{align}
with $\Omega_2$ found by the permutation $a_1\leftrightarrow a_2$.

\subsection{Thermodynamics and on-shell action}\label{subsec: thermo}

Let us now proceed to compute the rest of the thermodynamic properties of the solution. One can compute these objects from scratch, or alternatively obtain them in an $n\to\infty$ limit of corresponding quantities computed from the $(n;p_\alpha)$ solution; details of this latter route are provided in Appendix \ref{app: limits}.

In all, we will compute the following objects. Corresponding to the isometries of the $X_7$ background in which it lives, the black hole has an energy $E$, momentum $K$, and angular momenta $J_\alpha$. These correspond respectively to isometries along $t$, $\xi$ and $\phi_\alpha$. We also have a pair of electric charges $Q_I$, which in the 11-dimensional uplift correspond to independent rotations in $S^4$. The potentials that are thermodynamically dual to $(E,K,J_\alpha, Q_I)$ are denoted $(\beta, \Upsilon,\Omega_\alpha,\Phi_I)$. These are the inverse Hawking temperature $\beta=T^{-1}$ and angular velocities $(\Upsilon,\Omega_\alpha)$, which we already have, along with the electrostatic potentials $\Phi_I$. Finally, we have the Bekenstein-Hawking entropy $S$, and Euclidean on-shell action $I$.

The entropy $S$ is determined in terms of the area of the horizon\footnote{For simplicity of notation, in this section we assume $0\le a_1\le a_2\le g^{-1}$. However, all results are easily shown to hold for all $a_1,a_2\in (-g^{-1},g^{-1})$.},
\begin{align}
	S&=\fft{1}{4G_N}\int_{r=r_+,t} d^5x\,\sqrt{h}\,\Big|_{r=r_+}=\fft{1}{4G_N}\int_{a_1}^{a_2}dz\int_{a_2}^{g^{-1}}dy\int_0^{2\pi}d\xi \int_0^{2\pi} d\phi_1d\phi_2\,\sqrt{h}\,\Big|_{r=r_+}	\nn\\
	&= \frac{\pi^3 \lambda}{4 G_N \Xi_1 \Xi_2}\frac{\sqrt{\mathcal{S}(r_+)}}{r_+}
\label{eq:S}
\end{align}
Here, $h_{ij}$ is the five-dimensional induced metric on a constant $(t,r)$ slice, i.e.
\begin{align}
  h_{ij} = g_{\mu\nu}\frac{\partial x^\mu}{\partial x^i}\frac{\partial x^\nu}{\partial x^j} \bigg|_{t,r\,\text{const}}
\end{align}
Its determinant is given by
\begin{align}
  h= \frac{g^6 \lambda^2 y^2 z^2 (y^2 - z^2)^2}{(a_1^2 - a_2^2)^2 \Xi_1^4 \Xi_2^4}\left(\frac{g^2 \mathcal{S}(r)}{r^2} -  (r^2 + a_1^2)(r^2 + a_2^2)U(r)\right)
\end{align}
in terms of the function $\mathcal{S}$ defined in (\ref{eq: S def}). Next, the various momenta $\{K,J_1,J_2\}$ can be computed via Komar integrals over the constant $t$ slice at $r\to\infty$. For $K$ we find
\begin{align}
  K		&= -\frac{1}{16\pi G_N}\int_{r\to\infty,t} \star d (\partial_\xi)^\flat  \nn\\
 		&= \frac{\pi^2 \lambda^2 m }{16 G_N g \Xi_1 \Xi_2}\left((c_1+c_2)^2 -2g^2(s_1^2+s_2^2)(a_1-a_2)^2+g^4(c_1-c_2)^2(a_1-a_2)^4\right)
\end{align}
where here and throughout we use the standard musical notation $(\partial_\xi)^\flat $ to denote the 1-form canonically dual to the Killing vector\footnote{Note the opposite sign convention relative to \cite{Bobev:2023bxl}} $\partial_\xi$, i.e.
\begin{align}
	((\partial_\xi)^\flat)^\mu\partial_\mu=\partial_\xi\qquad\Longleftrightarrow\qquad (\partial_\xi)^\flat=g_{\mu\xi}dx^\mu\,.
\end{align}
Similarly, for the $J_\alpha$ we have
\begin{align}
  J_\alpha = - \frac{1}{16\pi G_N}\int \star d  (\partial_{\phi_\alpha})^\flat
\end{align}
We find then
\begin{align}
  J_1 &=-\fft{\pi^2\lambda m}{16G_N\Xi_1^2\Xi_2}\Bigg[4a_1c_1c_2+4(1-c_1c_2)(a_1-a_2)(1-a_1a_2g^2)\\
	&\quad+(c_1-c_2)^2\bigg(2a_2-a_1+g^2a_1(a_1^2 - a_2^2)\Big)\bigg)\bigg(1+2a_1a_2 g^2-g^2(a_1^2 + a_2^2)\bigg)\Bigg]
\end{align}
with $J_2$ found by the permutation $a_1\leftrightarrow a_2$.

The two electric charges $Q_I$ can also be computed by Komar integrals, as\footnote{Note the change in normalisation convention relative to \cite{Bobev:2023bxl}}
\begin{align}
  Q_I = -\frac{1}{16\pi G_N} \frac{1}{2g} \int \left(X_I^{-2} \star F^I_{(2)} - F^I_{(2)}\wedge A_{(3)}\right)
  \label{eq: QI def}
\end{align}
We then find
\begin{align}
  Q_1 =  \frac{\pi^2 \lambda m s_1}{8 G_N g \Xi_1 \Xi_2} \left(c_1 + c_2-g^2(c_1-c_2)(a_1-a_2)^2\right)
\end{align}
with $Q_2$ found by the permutation $\delta_1\leftrightarrow \delta_2$. Note that one in principle encounters ambiguities in defining conserved and quantised charges when, like here, the supergravity Lagrangian includes Chern-Simons terms \cite{Marolf:2000cb}. However for the solution in question, the fields die away sufficiently quickly that the Chern-Simons term in (\ref{eq: QI def}) does not contribute, and thus possibly different notions of charge coincide.

The electrostatic potentials are determined by
\begin{align}
  \Phi_I = l^\mu \!\!\left.\left( 2g A^I_{(1)\mu}\right)\right|_{r=r_+} - l^\mu \!\!\left.\left( 2g A^I_{(1)\mu}\right)\right|_{r\to\infty}
\end{align}
and thus the result does not depend on the pure gauge parameters $\alpha_I$. For $\Phi_1$, we find
\begin{align}
	\Phi_1=\frac{1}{g^2\mS(r_+)}\Bigg[& 2m r_+^2 \left(s_1(c_1+c_2)-g^2 s_1(c_1-c_2)(a_1-a_2)^2\right)\nn\\
	&\hspace{15mm} \times \left((1+g^2 r_+^2)(r_+^2 + a_1^2)(r_+^2 + a_2^2)+2m g^2 s_2^2(r_+^2 + a_1 a_2)\right)\nn\\
	&- 8m^2 g^2r_+^2 s_1 s_2^2(c_1-c_2)(a_1-a_2)^2(1-a_1 a_2 g^2) \Bigg]\,.
\label{eq:Phi}
\end{align}
with $\Phi_2$ determined by the permutation $\delta_1 \leftrightarrow \delta_2$. We should also impose that the norm of the gauge fields $A^I_{(1)}$ is regular at the horizon, which imposes
\begin{align}
  \alpha_I = - \Phi_I
\end{align}
We next use the ADM formalism for asymptotically locally AdS backgrounds to compute the energy $E$ \cite{Ashtekar:1984zz,Ashtekar:1999jx,Chen:2005zj}. First, one introduces a Weyl-rescaled metric $\tilde{g}_{ab} = \tilde{\Omega}^2 g_{ab}$ such that $\tilde{\Omega}=0$ but $d\tilde{\Omega}\neq 0 $ on the conformal boundary; the definition is insensitive to this choice, so let us choose $\tilde{\Omega}=1/gr$ for definiteness. Then, the energy is
\begin{align}
  E = \frac{1}{32\pi G_N g^3} \int_{r\to\infty,t} d^5 x f(y,z) \,\tilde{\Omega}^{-4} \tilde{n}^c \tilde{n}^d \tilde{C}^t{}_{cbd}(\partial_t)^b
  \label{eq: ADM energy}
\end{align}
The function $f(y,z)$ can be obtained from the induced 6-volume on the conformal boundary, and for the present solution is simply
\begin{align}
  f(y,z) = \frac{\lambda y z |y^2-z^2|}{g^2 \Xi_1^2 \Xi_2^2|a_1^2 - a_2^2|}
\end{align}
Meanwhile, $\bar n_{c} = \partial_{c}\widetilde{\Omega}$ and $C^{\mu}{}_{\nu\rho\sigma}$ is the Weyl tensor. Note that the Weyl tensor for $g_{\mu\nu}$ with the first index raised is equal to that for $\bar g_{\mu\nu}$, i.e., it is invariant under conformal rescaling of the metric.

Explicitly, we have $\tilde{n}^r = - g^3 r^2$ at leading order with all other components vanishing, and so ultimately we have
\begin{align}
  E = \frac{g^7}{32\pi G_N} \int_{r\to\infty,t} d^5 x f(y,z) \tilde{C}^t{}_{rtr}
\end{align}
We omit the explicit---and rather unweildy---expression for $\tilde{C}^t{}_{rtr}$, and simply state the result. Performing the integral, we find 
\begin{align}
  E=\frac{\pi^2 \lambda m }{16 G_N g \Xi_1^2 \Xi_2^2}&\left(2-g^2(a_1^2 + a_2^2)\right)\nn\\
  &\times\left((c_1+c_2)^2 -2g^2(s_1^2+s_2^2)(a_1-a_2)^2+g^4(c_1-c_2)^2(a_1-a_2)^4\right)
\end{align}
We then happily find that the quantities we have computed satisfy the first law of black hole thermodynamics,
\begin{align}
  dE = TdS + \Upsilon dK + \Omega_\alpha dJ_\alpha + \Phi_I dQ_I
\end{align}
where the variations are taken with respect to the parameters $(m,\lambda, a_\alpha \delta_I)$.

An important observable in Euclidean quantum gravity, particularly in the context of holography, is the Euclidean on-shell action. In lieu of a full computation via holographic renormalisation which we leave to future work, we can deduce the form of the Euclidean on-shell action for our solution by employing the so-called quantum statistical relation \cite{Gibbons:1976ue},
\begin{align}
  I(\beta,\Upsilon,\Omega_\alpha,\Phi_I) 	&= -S+ \beta\left(E - \Upsilon K - \Omega_\alpha J_\alpha - \Phi_I Q_I \right)	
  \label{eq: OSA}
\end{align}
where we have computed everything on the right-hand-side, and we view $\{S,K,J_\alpha,Q_I\}$ implicitly as functions of $\{\beta,\Upsilon,\Omega_\alpha,\Phi_I\}$.

\subsection{Supersymmetric limits}\label{subsec: SUSY limits}

We would next like to determine and study regimes of parameter space for which our solution preserves some supersymmetry. A thorough investigation of this question requires a careful analysis of the supersymmetry variations of the supergravity theory, and the construction of the corresponding Killing spinors. Unfortunately this is a technically difficult task. However, there is alternative and slightly less direct route one often takes when analysing the supersymmetric properties of black holes in supergravity; see for instance \cite{Cvetic:2005zi} for a detailed discussion. Briefly, for each pair of conjugate supercharges preserved by the background, one can infer a BPS bound obeyed by some linear combination of the mass $E$, the momenta $J_\alpha, K$, and electric charges $Q_I$. Then, a solution is annihilated by a given pair of supercharges precisely if the corresponding BPS bound is saturated.

The $X_7 \times S^4$ background preserves $24$ real supercharges, leading to 12 BPS bounds. Four of these are degenerate, imposing simply $K\ge 0$. The remaining $8$ are
\begin{align}
  E+ J_1 + J_2 \pm 2Q_1 \pm 2Q_2 \ge 0,\qquad E- J_1 - J_2 \pm 2Q_1 \pm 2Q_2 \ge 0
\end{align}
Without loss of generality we focus on
\begin{align}
  E-J_1 - J_2 - 2Q_1 - 2Q_2 \ge 0
  \label{eq: bulk BPS bound}
\end{align}
Then, by inspection we find that our black hole solution saturates this bound precisely when its parameters satisfy
\begin{equation}
	(a_1g+a_2g-1)=\fft{2}{1-e^{\delta_1+\delta_2}}\,.\label{susy}
\end{equation}
Thus, we find a five-parameter family of regular supersymmetric Euclidean black hole solutions. For generic parameter values satisfying (\ref{susy}), we expect that the solution preserves two real supercharges\footnote{There may be special further subspaces of parameter space for which two or more BPS bounds degenerate and the solution preserves additional supercharges. We do not explore this possibility here.}.

Working instead in the grand canonical ensemble, the constraint (\ref{susy}) implies a linear relation,\footnote{The choice of sign here corresponds to a pair of conjugate solutions for the parameter $m$ as a function of $r_+$.}
\begin{align}
  \beta \left[2+\sum_\alpha \Omega_\alpha - \sum_I \Phi_I\right] = \pm 2\pi i
  \label{eq: potentials linear relation}
\end{align}
Note, we are working with a Euclidean solution, and so there is nothing wrong with this being complex; see \cite{Bobev:2023bxl} and references therein for further discussion. 

Remarkably, when on the supersymmetric locus (\ref{susy}), or equivalently when grand canonical parameters are constrained as in (\ref{eq: potentials linear relation}), the form of the on-shell action $I$ as defined in (\ref{eq: OSA}) simplifies dramatically. We find simply
\begin{align}
  I(\beta,\Upsilon,\Omega_\alpha,\Phi_I) = -\frac{\pi^2}{128 G_N g^5}\frac{\varphi_1^2 \varphi_2^2}{\beta\Upsilon \omega_1 \omega_2}
  \label{eq: SUSY OSA}
\end{align}
where we define
\begin{align}
  \omega_\alpha = \beta(1-\Omega_\alpha),\qquad \varphi_I = \beta(2-\Phi_I)
  \label{eq: curly variables}
\end{align}
which by virtue of (\ref{eq: potentials linear relation}) satisfy
\begin{align}
  \omega_1 + \omega_2 - \varphi_1 - \varphi_2 = \mp 2\pi i
\end{align}
for supersymmetric black holes.\\

Let us finally discuss Lorentzian supersymmetric solutions. For generic choices of parameters satisfying (\ref{susy}), the Euclidean supersymmetric solution does not admit a well-defined analytic continuation to Lorentzian signature. By this we mean that in analytically continuing, we necessarily introduce some pathology, either rather manifestly through complex fields, or more subtly through causal issues like closed timelike curves\footnote{To see this, we rely on the fact that at all finite $n$, the solution exhibits such pathologies \cite{Bobev:2023bxl}. In principle, one could imagine these pathologies going away in the strict $n\to\infty$ limit; we do not exlore this possibility here. Similarly, the well-behavedness of the extremal supersymmetric solution is assumed by that of the finite $n$ solution.}. 

However, by imposing an additional constraint, such issues can be avoided and a good Lorenztian solution obtained. In the grand canonical ensemble, this constraint is precisely that of extremality,
\begin{align}
  T=0
\end{align}
We thus end up with a four-parameter family of good Lorentzian extremal supersymmetric solutions, which in keeping with much of the literature, we dub \textit{BPS}. A BPS black hole is then specified by parameters $\{\lambda,a_\alpha,\delta_I\}$ satisfying (\ref{susy}), with the final parameter $m$ fixed as
\begin{align}
  m= m^\star = \frac{M_1}{g^4 M_2}
\end{align}
 where
 \begin{align}
  M_1	&= 8g\Xi_1^2\Xi_2^2(a_1+a_2)\left(-2+3g(a_1+a_2)-2g^2(a_1^2 +a_2^2 +a_1 a_2)+g^3 a_1 a_2(a_1 +a_2)\right)	\nn\\
  M_2	&= (2-g(a_1+a_2))^2 \bigg[ \,\,4(2-g(a_1+a_2))		\nn\\
  		&\hspace{5mm} + 2(s_1^2 + s_2^2) \left(-2+5g(a_1 + a_2)-2g^2(a_1^2 + a_2^2 + 4a_1 a_2)+ g^3(a_1+a_2)(a_1^2 + a_2^2) \right)\nn\\
  		&\hspace{5mm} - (c_1-c_2)^2(2-g(a_1+a_2))\left(1+4g^2(a_1-a_2)^2 - g^4(a_1-a_2)^2(a_1+a_2)^2\right) \bigg]
\end{align}
Here and in what follows we use a ``$\star$'' superscript to denote the BPS value of an object.

The horizon radius of these BPS solutions takes a simple closed form,
\begin{align}
  r_+^\star = \frac{1}{g}\sqrt{\frac{g \left(a_1+a_2 - 2 a_1 a_2g\right)}{(a_1+a_2)g - 2}}
\end{align}
while various potentials take the BPS values
\begin{align}
  \Upsilon^\star = 0,\quad \Omega_\alpha^\star = 1,\quad \Phi_I^\star = 2
\end{align}
but note importantly that the combinations $\beta\Upsilon$ and $\omega_\alpha,\varphi_I$ as defined in (\ref{eq: curly variables}) remain finite and generically non-zero in the BPS, and thus the on-shell action retains its functional form (\ref{eq: SUSY OSA}). 

Working instead in the microcanonical ensemble, while the supersymmetry constraint (\ref{susy}) implies that $E$ saturates the BPS bound (\ref{eq: bulk BPS bound}), the additional constraint $T=0$ manifests as a particular non-linear constraint amongst the BPS charges $\{K^\star,J^\star_\alpha, Q^\star_I\}$, given by
\begin{align}
  &\frac{Q_1^\star Q_2^\star(Q_1^\star+Q_2^\star)-\frac{\pi^2}{16G_N g^5}K^\star(J_1^\star+J_2^\star)}{Q_1^\star+Q_2^\star}			\nn\\
  &\quad =\left(\frac{1}{2}\left((Q_1^\star)^2+ (Q_2^\star)^2\right)+2Q_1^\star Q_2^\star+\frac{\pi^2}{16G_N g^5}K^\star\right)	\nn\\
  &\hspace{10mm}\times\left(1-\sqrt{1-\frac{(Q_1^\star)^2 (Q_2^\star)^2+\frac{\pi^2}{8 G_N g^5}K^\star J_1^\star J_2^\star}{(\frac{1}{2}\left((Q_1^\star)^2+ (Q_2^\star)^2\right)+2Q_1^\star Q_2^\star+\frac{\pi^2}{16G_N g^5}K^\star)^2}}\right)  
  \label{eq: non-linear rel}
\end{align}
Similar non-linear constraints are indeed ubiquitous in BPS black hole solutions; see for instance \cite{Cassani:2019mms} for a helpful review.

Finally, note that the entropy of the BPS solutions takes a compact form,
\begin{align}
  S^\star(K^\star,J_\alpha^\star ,Q_I^\star) = 2\pi \sqrt{\frac{Q_1^\star Q_2^\star(Q_1^\star+Q_2^\star)-\frac{\pi^2}{16G_N g^5}K^\star(J_1^\star+J_2^\star)}{Q_1^\star+Q_2^\star}}
  \label{eq: BPS entropy}
\end{align}

\section{Holography}
\label{sec: holography}

With a solution and its thermodynamics in hand, let us finally discuss what these results mean for the dual superconformal quantum mechanics.

\subsection{The dual theory, briefly}

The dual theory is a supersymmetric non-relativistic conformal field theory. It has as its spacetime symmetries the Schr\"odinger group Schr$(4)$, corresponding to the isometries of $X_7$. The pp-wave background at the conformal boundary of $X_7$ corresponds to an analogue of radial quantisation in the NRCFT, in which local operators are mapped to states in a harmonic trap \cite{Nishida:2007pj}. Many more details on this non-relativistic operator-state map in this context can be found in \cite{Dorey:2023jfw}\footnote{Note that what we call $E$ here is called $\Delta$ there.}. All we need to know here is that we can organise states by their charges under the Cartan subalgebra, generated by the oscillator Hamiltonian $E$, the $SO(4)\subset$ Schr$(4)$ commuting rotations $J_\alpha$, and the particle number $K$. The theory also has an $SO(5)$ R-symmetry, corresponding to the isometries of $S^4$ in the bulk. We take R-symmetry Cartan generators $Q_I$. Finally, the theory has 24 real supercharges.

In a sector of fixed $K$, the theory is described by a non-linear $\sigma$-model on $\M_{K,N}$, the moduli space of $K$ Yang-Mills instantons in $SU(N)$. $\M_{K,N}$ has complex dimension $4KN$. This manifold is a hyper-K\"ahler cone, which ensures that it admits maximal superconformal symmetry $\frak{osp}(4^*|4)$ \cite{Singleton:2016hky}. There is an additional $SU(2)$ global symmetry, along with additional non-linearly realised symmetries, which together make up precisely the supersymmetric extension of Schr$(4)\times SO(5)$ described above \cite{Aharony:1997an}. There is however also an $SU(N)$ global symmetry, which can be traced back to the global part of the gauge group of the $U(N)$ $(2,0)$ theory. The top-down construction of the duality instructs us to restrict our attention to singlet states under this $SU(N)$, which we do.

There are subtleties however in defining this SCQM precisely, due to small-instanton singularities in $\M_{K,N}$. There are essentially two ways to define it concretely. One is to stick with the non-linear $\sigma$-model, but replace $\M_{K,N}$ with a resolved space $\widetilde{\M}_{K,N}$ \cite{Dorey:2018klg,Singleton:2016hky,Barns-Graham:2018qvx,Barns-Graham:2018xdd}. An alternative and perhaps more straightforward route is to work in the Berkooz-Douglas matrix model for the D0--D4 system, which is perfectly well-defined \cite{Berkooz:1996is}. This is nothing but the $U(K)$ BFSS model \cite{Banks:1996vh}, to which we have added $N$ fundamental hypermultiplets. Then, in the limit $g_\text{YM}\to\infty$, the theory reduces to a non-linear $\sigma$-model on its Higgs branch $\M_{K,N}$, i.e. the SCQM we wish to study. Thus, we can \textit{define} our theory as the strongly-coupled fixed point of this matrix model.\\

Our notation makes the dictionary between symmetries on either side of the duality entirely manifest. All that remains is to state the relationship between the parameter $N$ of the SCQM, and the radius $R_\text{AdS}=g^{-1}$, as well as the radius $R_{S^4}$ of the $S^4$. We have
\begin{align}
  \frac{R_\text{AdS}}{l_p} = \frac{2R_{S^4}}{l_p} = 2( \pi N)^{1/3}
  \label{eq: AdS radius dictionary}
\end{align}
in terms of the 11-dimensional Planck length $l_p$. In our conventions, we have 11-dimensional Newton's constant $G_N^{(11)}=16\pi^7l_p^9$, giving rise to 7-dimensional Newton's constant $G_N^{(7)}=\text{Vol}(S^4)^{-1}G_N^{(11)} = 6\pi^5 (\pi N)^{-4/3}l_p^5$. We thus have
\begin{align}
  \frac{R_\text{AdS}^5}{G_N^{(7)}} = \frac{1}{g^5 G_N^{(7)}} = \frac{16 N^3}{3\pi^2}
  \label{eq: dictionary}
\end{align}
Everywhere else in the text we write $G_N = G_N^{(7)}$.

\subsection{The supergravity regime}

As in any holographic duality, we need to understand in which regime of parameter space we can trust classical supergravity in the bulk. That is, for what values of parameters can our black hole solution be trusted as an M-theory background. This question is somewhat more subtle than in conventional AdS/CFT setups, and was first addressed in \cite{Maldacena:2008wh}.

Firstly, we should certainly have that the AdS radius $R_\text{AdS}=g^{-1}$ is large in Planck units, so as to suppress derivative corrections to the supergravity action; from (\ref{eq: AdS radius dictionary}), we have $N\gg 1$. But we are not done yet: recall that the asymptotic geometry $X_7$ contains a null circle. M-theory formulated on such a geometry contains additional light states wrapping this null circle that do not decouple at low energies; the correct low energy description is given not in terms of supergravity, but D0-brane quantum mechanics \cite{Dine:1997sz,Seiberg:1997ad}. However, as first pointed out in \cite{Maldacena:2008wh}, we can do better for the black hole solution. In particular, there are regions of parameter space for which the null circle in $X_7$ in fact becomes spacelike and very large in Planck units, up to some characteristic radius around the black hole. Thus, in such a regime, we can trust our black hole solution as a good M-theory background.

Let us determine the requisite region of parameter space. We compute the invariant length-squared of the circle in Planck units,
\begin{align}
 &\frac{g(\partial_\xi,\partial_\xi)}{l_p^2} \nn\\
 &= 2\pi^{2/3}(\lambda N^{1/3})^2\frac{m(1-g^2 y^2)^2(1-g^2 z^2)^2}{\Xi_1^2 \Xi_2^2(r^2+y^2)^2(r^2+z^2)^2}	\Bigg( -8g^2 m(a_1-a_2)^2 s_1^2 s_2^2		\\
 &\hspace{5mm} +\left((c_1+c_2)^2-2g^2(a_1-a_2)^2(s_1^2+s_2^2)+g^4(a_1-a_2)^4(c_1-c_2)^2\right)(r^2 + y^2)(r^2 +z^2) \Bigg)\nn 
\end{align}
We can consider in particular this length evaluated at the horizon, $r=r_+$. Note then that $r_+$ takes the form
$r_+ = g^{-1}f(g^4 m,ga_\alpha)$ for a function $f$ we leave implicit. It follows then that a sufficient\footnote{This is of course not the only region of parameter space in which $(g_{\xi\xi}/l_p^2)\gg 1$ at the horizon, but just in a sense the simplest. We leave the study of other regions to future work.} condition on the parameters $(m,\lambda,a_\alpha,\delta_I)$ for the circle to be large in Planck units at the horizon is\footnote{There is in fact a small portion of the horizon where the circle remains of Planckian length regardless of how we scale $\lambda$ and $N$, as first noted in \cite{Maldacena:2008wh}. This is when $1-(\lambda N^{1/3})^{-1}<g^2 y^2 < 1$, which corresponds to the region near $\partial(\R^4)=S^3_\infty$ in the $S^1 \times \R^4$ horizon. However, the contribution $S_\infty$ of this region to the entropy $A=A/4G_N$ as a proportion of the total entropy (\ref{eq:S}) scales a $S_\infty/S\sim (\lambda N^{1/3})^{-1}$, and is thus negligible in the regime of interest. So, like the authors of \cite{Maldacena:2008wh}, we assume we can neglect this region.  }
\begin{align}
  \lambda N^{1/3} \gg 1 
\end{align}
while we hold the dimensionless combinations $\{g^4m,ga_\alpha,\delta_I\}$ fixed. We can then continue to trust our solution from the horizon at $r_+\sim R_\text{AdS}$ all the way out to a parametrically larger radius $r_\text{max}\sim (\lambda N^{1/3})^{1/4} R_\text{AdS}$, where the circle shrinks to a size comparable to $l_p$.\\

In summary, we can trust our supergravity solution in the regime
\begin{align}
  \lambda N^{1/3}, N \gg 1,\qquad \text{with }\{g^4m, ga_\alpha,\delta_I\} \text{ held fixed}
  \label{eq: SUGRA regime, params}
\end{align}
But $\lambda$ is just a parameter of our black hole solution. We need to translate this regime to something physically meaningful. To do so, note first that we have the following scaling with $\lambda$ and $N$,
\begin{align}
  \Upsilon \sim \lambda^{-1} ,\qquad \beta,\Omega_\alpha,\Phi_I\sim 1,\qquad K\sim \lambda^2 N^3 ,\qquad E,J_\alpha,Q_I\sim \lambda N^3
\end{align}
Therefore, working in the grand canonical ensemble, the regime (\ref{eq: SUGRA regime, params}) corresponds to sitting in a regime with
\begin{align}
  N\gg 1,\quad \Upsilon \ll N^{1/3}, \quad \text{with }\beta,\Omega_\alpha, \Phi_I \sim 1
\end{align}
Working instead microcanonically, the relevant regime is
\begin{align}
   K \gg N^{7/3} \gg 1,\quad \text{with } E,J_\alpha, Q_I \sim K^{1/2}N^{3/2}
   \label{eq: sugra regime microcanonical}
\end{align}
In particular, in this regime the entropy scales as
\begin{align}
  S\sim K^{1/2}N^{3/2}
\end{align}
Let us finally make a more speculative comment about alternative formulations of the bulk theory. It is important to appreciate that we are in a very different scenario compared to conventional AdS/CFT. Usually, in a suitable regime of parameters the conformal vacuum of the CFT certainly admits a realisation in classical gravity, as pure AdS. Step two is to then consider black hole solutions. Here, we have something much weirder: while we can trust our black hole solution in the bulk in the regime (\ref{eq: sugra regime microcanonical}), what we would like to call the vacuum state, i.e. pure $X_7$, is never a good M-theory background for any value of parameters! Seemingly, the vacuum of our NRCFT does not admit a realisation in any classical theory of gravity.

One can try to remedy this uncomfortable situation by exploring alternative formulations of the bulk theory. A simple step is to reduce to Type IIA when the circle becomes small; in doing so, we are able to extend the regime of validity of our black hole to $K\gg N \gg 1$ \cite{Maldacena:2008wh}. However, this does not solve our vacuum problem. A more adventurous possibility is to use certain proposed string dualities, which relate null-compactified string/M-theory to non-relativistic string/M-theory. In detail, by first reducing null-compactified M-theory on an auxiliary circle to IIA, performing a double T-duality \cite{Bergshoeff:2018yvt,Lahnsteiner:2022mag} to non-relativistic IIA \cite{Oling:2022fft}, and then lifting back up to 11-dimensions, we arrive at the striking idea that M-theory on $X_7\times S^4$ may admit a dual description in the language of `non-relativistic M-theory' \cite{Blair:2021waq}. In particular, in a suitable regime of parameters, one expects the vacuum of our NRCFT to correspond to some vacuum state in the low energy limit of this bulk theory, known as membrane Newton-Cartan gravity. We leave an investigation of such a connection to future work.

\subsection{Relating partition functions}

Next, we would like to work out in detail the holographic correspondence in its most primal form: `$\ZZ=\ZZ$'.

In the bulk, the partition function $\ZZ_\text{grav}(\beta,\Upsilon,\Omega_\alpha,\Phi_I)$ can at least formally be defined as the Euclidean path integral with $X_7$ boundary conditions at $r\to\infty$, where the thermal circle is twisted as
\begin{align}
  (\tau,\xi,\phi_\alpha) \sim (\tau+\beta,\xi+i\beta\Upsilon,\phi_\alpha+i\beta \Omega_\alpha)
  \label{eq: bulk thermal circle}
\end{align}
and we turn on a flat connection VEV for the gauge fields,
\begin{align}
  A_I \xrightarrow{r\to\infty}   i\Phi_I d\tau
  \label{eq: non-norm mode}
\end{align}
This flat connection can equivalently be absorbed into matter fields, at the expense of introducing a twist along R-symmetry orbits as we traverse the thermal circle.

In the supergravity regime $K \gg N^{7/3}\gg 1$, this partition function is dominated by our Euclidean black hole solution; neglecting fluctuations around this classical solution, we have
\begin{align}
  -\log \ZZ_\text{grav}(\beta,\Upsilon,\Omega_\alpha,\Phi_I) \sim I(\beta,\Upsilon,\Omega_\alpha,Q_I)
\end{align}
with $I(\beta,\Upsilon,\Omega_\alpha,Q_I)$ as given in (\ref{eq: OSA}).\\

Conversely, in the dual theory we can define a partition function
\begin{align}
  \ZZ_{\text{CFT}}(\mu,\tilde{\beta},w_\alpha,u_I) = \text{Tr}_{\mathcal{H}}\,e^{-\mu K - \tilde{\beta} E - w_\alpha J_\alpha - u_I Q_I }
\end{align}
as a trace over the Hilbert space. Decomposing, we have
\begin{align}
  \ZZ_{\text{CFT}}(\mu,\tilde{\beta},w_\alpha,u_I) = \sum_{K=0}^\infty e^{-\mu K} \ZZ_{\text{QM}}^{(K)}(\tilde{\beta},w_\alpha,u_I)
  \label{eq: sum over QM pfs}
\end{align}
where $\ZZ_{\text{QM}}^{(K)}$ is a partition function of the superconformal quantum mechanics on $\M_{K,N}$, defined as
\begin{align}
  \ZZ_{\text{QM}}^{(K)}(\tilde{\beta},w_\alpha,u_I) = \text{Tr}_{\mathcal{M}_{K,N}}^{\text{singlets}}\,e^{- \tilde{\beta} E - w_\alpha J_\alpha - u_I Q_I }
\end{align}
where the trace is taken only over $SU(N)$ singlets.\\

We would like to relate $\ZZ_\text{grav}$ to $\ZZ_\text{CFT}$. To do so, we need to relate the two sets of arguments $\{\beta,\Upsilon,\Omega_\alpha, \Phi_I\}$ and $\{\mu,\tilde{\beta}, w_\alpha, u_I\}$. As usual, this is achieved by regarding $\ZZ_{\text{CFT}}$ as a Euclidean path integral. For each $K$, we can formulate $\ZZ_{\text{QM}}^{(K)}$ in the usual way as a Euclidean path integral in the quantum mechanics on $\M_{K,N}$. Alternatively, in the NRCFT as a whole, we can formally write down a Euclidean path integral expression for $\ZZ_{\text{CFT}}$, where the relevant action was derived and studied in \cite{Lambert:2019nti,Mouland:2019zjr}. Here however we need here a third perspective. Namely, that $\ZZ_{\text{CFT}}$ can be regarded as the Euclidean partition function of the six-dimensional $U(N)$ $(2,0)$ superconformal field theory \cite{Aharony:1997an,Dorey:2023jfw}, formulated on a compactified pp-wave geometry ($\xi\sim \xi+2\pi$) with thermal circle
\begin{align}
  (\tau,\xi,\phi_\alpha) \sim (\tau + \tilde{\beta},\xi- i\mu,\phi_\alpha - iw_\alpha)
  \label{eq: boundary thermal circle}
\end{align}
We must also turn on a background R-symmetry flat connection,
\begin{align}
  \tilde{A}_I = - i\frac{u_I}{\tilde{\beta}}d\tau 
\end{align}
Now we're good to go. First, we compare the thermal circles (\ref{eq: bulk thermal circle}) and (\ref{eq: boundary thermal circle}). We immediately identify inverse temperatures simply as
\begin{align}
  \tilde{\beta} = \beta
\end{align}
and so just drop the tilde from now on. Otherwise, we have
\begin{align}
  \mu = -\beta \Upsilon,\qquad w_\alpha = -\beta \Omega_\alpha\ ,
\end{align}
Finally, the non-normalisable mode (\ref{eq: non-norm mode}) of $A_I$ fixes the VEV of $\tilde{A}_I$, from which we read off
\begin{align}
   u_I = - \beta \Phi_I 
\end{align}
In summary, using the regularity of the bulk solution, we have arrived at the identification
\begin{align}
  \ZZ_\text{CFT}(\mu,\beta,w_\alpha,u_I) = \ZZ_\text{grav}\left(\beta, \Upsilon=-\frac{\mu}{\beta},\Omega_\alpha = - \frac{w_\alpha}{\beta},\Phi_I = - \frac{u_I}{\beta}\right)
  \label{eq: Z=Z}
\end{align}
Then, in terms of the variables $\{\mu,\beta,w_\alpha,u_I\}$, we can trust the supergravity solution so long as $N\gg 1$ and $\mu \ll N^{1/3}$, with $\beta,w_\alpha,u_I=\mathcal{O}(1)$. In particular, at leading order in a large $N$ expansion, we should find
\begin{align}
  -\log \ZZ_\text{CFT}(\mu,\beta,w_\alpha,u_I) \sim  I\!\left(\beta, \Upsilon=-\frac{\mu}{\beta},\Omega_\alpha = - \frac{w_\alpha}{\beta},\Phi_I = - \frac{u_I}{\beta}\right)
  \label{eq: Z=Z leading order}
\end{align}
in terms of the on-shell action $I$.

\subsection{Matching the quantum mechanics}

We would like to independently verify (\ref{eq: Z=Z leading order}) from the superconformal quantum mechanics. To make progress, we specialise to a codimension-$1$ subspace of parameters,
\begin{align}
  2\beta - w_1-w_2+u_1+u_2 = 2\pi i\quad (\text{mod }4\pi i)
\end{align}
On this subspace, $\ZZ_\text{CFT}$ becomes a maximally-refined superconformal index. Let us write $\I(\mu,\beta,w_\alpha,u_I)=\ZZ_\text{CFT}(\mu,\beta,w_\alpha,u_I)$ on this subspace, which is in turn a sum as in (\ref{eq: sum over QM pfs}) over superconformal indices of the superconformal quantum mechanics on each $\M_{K,N}$, as formalised in \cite{Dorey:2018klg} (see also \cite{Kim:2011mv}). Then, $\I$ counts only states annihilated by a certain supercharge $\mathcal{Q}$, which carries charges $(0,+1,-\frac{1}{2},-\frac{1}{2},-\frac{1}{2},+\frac{1}{2},+\frac{1}{2})$ under $(K,E,J_1,J_2,J_2,Q_1,Q_2)$. Thus, the states counted by $\I$ saturate the BPS bound
\begin{align}
  \{\mathcal{Q},\mathcal{S}\} = E-J_1-J_2 - 2Q_1 - 2Q_2,\qquad \mathcal{S} = \mathcal{Q}^\dagger 
\end{align}
that we saw already in the bulk in (\ref{eq: bulk BPS bound}). 

It is useful to perform the recasting
\begin{align}
  \I(\mu,\beta,w_\alpha,u_I) 	&=  \text{Tr}_{\mathcal{H}}\,e^{-\mu K - \beta E - w_\alpha J_\alpha - u_I Q_I }	\nn\\
  								&=  \text{Tr}_{\mathcal{H}}\,e^{\beta\{\mathcal{Q},\mathcal{S}\}-\mu K - \beta E - w_\alpha J_\alpha - u_I Q_I }	\nn\\
  								&=   \text{Tr}_{\mathcal{H}}  \exp\bigg[ -\mu K- \underbrace{(w_\alpha + \beta )}_{\begin{array}{c}
  \omega_\alpha
\end{array}
} J_\alpha - \underbrace{(u_I+2\beta)}_{
\begin{array}{c}
	\varphi_I
\end{array}
} Q_I\bigg]\,,
\label{eq: index recasting}
\end{align}
where under the dictionary (\ref{eq: Z=Z}) we identify the parameters $\omega_\alpha = w_\alpha + \beta = \beta(1-\Omega_\alpha)$ and $\varphi_I = u_I + 2\beta = \beta(2-\Phi_I)$ that we defined already in (\ref{eq: curly variables}). Thus, we can view $\I=\I(\mu,\omega_\alpha,\varphi_I)$ where
\begin{align}
  \omega_1 + \omega_2 - \varphi_1 - \varphi_2 = 2\pi i \quad (\text{mod }4 \pi i)
  \label{eq: omega phi constraint}
\end{align}
This constraint precisely ensures that the Euclidean black hole solution that dominates $\I$ is supersymmetric. Then, using the compact form (\ref{eq: SUSY OSA}) of the supersymmetric on-shell action, as well as the relation (\ref{eq: dictionary}), we arrive at the prediction
\begin{align}
  -\log \I(\mu,\omega_\alpha,\varphi_I)  = \frac{N^3}{24} \frac{\varphi_1^2 \varphi_2^2}{\mu \omega_1 \omega_2}
  \label{eq: index asymptotic}
\end{align}
for the leading order growth of the index in the supergravity regime.\\

One would then like to explicitly recover (\ref{eq: index asymptotic}) independently in the microscopic theory. Precisely this was achieved in \cite{Dorey:2022cfn}. In detail, this work studied the relevant asymptotics of the superconformal index of the superconformal quantum mechanics on $\M_{K,N}$, which after a change of ensemble gives rise to the index $\I$ as in (\ref{eq: sum over QM pfs}). Then, precisely the asymptotic (\ref{eq: index asymptotic}) was recovered at leading order as $\mu\to 0$ with $N\gg 1$. It was additionally shown that one can also recover these same asymptotics in the alternative, overlapping Cardy-like\footnote{One should be careful here, as there are in fact several Cardy-like limits one may consider, depending on which sheet of the index we are on \cite{Cassani:2021fyv}. The limit described here is naturally regarded as the index on the second sheet in analogy with \cite{Cassani:2021fyv}, as the form of (\ref{eq: index recasting}) along with the condition (\ref{eq: omega phi constraint}) means that fermionic boundary conditions along the thermal circle are twisted by R-symmetry charges. Conversely, the Cardy-like limit on the first sheet corresponds for instance to $\omega_1\to 2\pi i, \omega_2\to 0$, crucially while always maintaining (\ref{eq: omega phi constraint}).} regime of $\mu\to 0$ and $\omega_1,\omega_2\to 0$, valid at all $N$.

Let us emphasise that the result (\ref{eq: index asymptotic}) has been recovered purely by studying the superconformal quantum mechanics on $\M_{K,N}$ and its index. Nonetheless, let us briefly comment on ways in which one might arrive at this result starting with the superconformal index of the $U(N)$ six-dimensional $(2,0)$ theory. One must start with an index computed on $S^1\times (S^5/\Z_n)$. It was then shown in \cite{Dorey:2022cfn} that the explicit expression for this lens space index as presented in \cite{Kim:2013nva} goes over precisely to the quantum mechanical index $\I$ in the limit dual to the bulk limit in Section \ref{subsec: US limit}, in which $n\to \infty$ while a certain chemical potential goes to zero like $1/n$. From here, we can simply follow the methods of \cite{Dorey:2022cfn} to extract out (\ref{eq: index asymptotic}), as discussed above. There is however another route. We can first consider a Cardy-like limit on the second sheet of the six-dimensional index, in which a description in terms of an effective field theory on $S^5/\Z_n$ emerges\footnote{To our knowledge, this computation has only been performed for $n=1$, which would need to be generalised to all $n$ for this computation.} \cite{Choi:2018hmj} (see also \cite{Nahmgoong:2019hko,Lee:2020rns}), and a simple expression for its asymptotics is obtained. \textit{Now} taking $n\to\infty$ with suitable scaling of chemical potentials, one expects to once again land on (\ref{eq: index asymptotic}), where the six-dimensional second sheet Cardy-like limit implies that we are in the regime $\omega_1,\omega_2\to 0$. We do not work out the details of this computation here.

\subsection{Hilbert space perspective}

We would finally like to say something about the Lorenztian perspective on the holographic match, and thus say something about black hole microstates.

Let us first simply state an empirical fact. Recall that we found a four-parameter family of Lorentzian black hole solutions that were both supersymmetric and extremal, which we dubbed BPS. Their energy $E^\star$ is determined by the BPS bound (\ref{eq: bulk BPS bound}), while their remaining charges $\{K^\star,J_\alpha^\star,Q_I^\star\}$ obey the non-linear relation (\ref{eq: non-linear rel}). Their entropy $S^\star$ is given compactly in (\ref{eq: BPS entropy}), or after using (\ref{eq: dictionary}), as (\ref{eq: intro S}). We then find that this very same entropy can also be found as
\begin{align}
  S^\star(K^\star,J_\alpha^\star,Q_I^\star) = \text{ext}_{\{\mu,\omega_\alpha,\varphi_I\}} \left[- \frac{N^3}{24} \frac{\varphi_1^2 \varphi_2^2}{\mu \omega_1 \omega_2} + \mu K^\star + \omega_\alpha J_\alpha^\star + \varphi_I Q_I^\star\right]
  \label{eq: ext principle}
\end{align}
where the extremisation is performed over the variables $\{\mu,\omega_\alpha,\varphi_I\}$, subject to the constraint (\ref{eq: omega phi constraint}). This result was essentially derived in \cite{Dorey:2022cfn}. Formulae such as these for BPS entropy have appeared in the literature now for some time, referred to usually as an \textit{extremisation principle} \cite{Hosseini:2017mds,Hosseini:2018dob}. Let us explain what it means.\\

Let us go right the way back and consider once again our index $\I$ in the form written in (\ref{eq: index recasting}). We can without loss of generality assume that $\omega_1 + \omega_2 - \varphi_1 - \varphi_2 = 2\pi i$. Then, it is convenient to trade $\{J_\alpha,Q_I\}$ for a new basis of linearly independent charges $\{P_a,F\}$, $a=1,2,3$, with the following properties\footnote{A standard satisfying these requirements is $F=2Q_2, P_1 = J_1+\frac{1}{2}(Q_1+Q_2),P_2 = J_2+\frac{1}{2}(Q_1+Q_2),P_3=Q_1-Q_2$. But everything we say holds for all choices.}. Each of the $P_a$ commutes with the supercharge used to define the index, $[P_a,\mathcal{Q}]=0$, while the final element has $[F,\mathcal{Q}]=\mathcal{Q}$, and thus can be regarded as a fermion number\footnote{One usually also requires $F$ to have integer eigenvalues, but we won't need to assume that here.}. It follows then that we can write
\begin{align}
  -\omega_\alpha J_\alpha  - \varphi_I Q_I = i \pi F - b_a P_a
\end{align}
for three new independent chemical potentials $b_a$, and thus the index takes the form
\begin{align}
  \I(\mu,b_a) 	&= \text{Tr}_\mathcal{H}\, e^{i\pi F} e^{-\mu K - b_a P_a}		\nn\\
  				&= \sum_{\text{BPS states}} d_\text{BPS}(F,K,P_a)e^{i\pi F}e^{-\mu K - b_a P_a} 
  				\label{eq: index recast again}
\end{align}
where $d_\text{BPS}(F,K,P_a)$ denotes the degeneracy of BPS states with charges $\{F,K,P_a\}$.\\

Holography then dictates that at leading order in the supergravity regime, we should find a relation of the form
\begin{align}
  d_\text{BPS}(F,K,P_a) \sim e^{S_\text{BPS}(F,K,P_a)}
  \label{eq: central dogma}
\end{align}
where $S_\text{BPS}(F,K,P_a)$ is the Bekenstein-Hawking entropy of some Lorentzian supersymmetric black hole with charges $\{F,K,P_a\}$, which recall are just some linear combinations of $\{K,J_\alpha, Q_I\}$.\\

There are then issues in accessing both sides of (\ref{eq: central dogma}). Looking first at the right-hand-side, recall that we only found a \textit{four} parameter family of well-behaved Lorentzian supersymmetric black hole solutions, which are also extremal. For the purpose of the present discussion, it is useful to cast the corresponding constraint (\ref{eq: non-linear rel}) as a determination of $F= F_\star (K,P_a)$ as a function of $\{K,P_a\}$. Conversely, looking now at the left-hand-side, we do not have a way to access the BPS degeneracies $d_\text{BPS}$. The best we can do is extract the coefficients in the $e^{-\mu K}e^{-b_a P_a}$ expansion of the index (\ref{eq: index recast again}), which are
\begin{align}
  \mathcal{C}(K,P_a):= \sum_{F} e^{i\pi F} d_\text{BPS}(F,K,P_a)
\end{align}
These coefficients are then computed by a Legendre transform that is trivially related to the one we already did in (\ref{eq: ext principle}).

Then, the extremisation principle  (\ref{eq: ext principle}) is precisely restated as follows. At leading order as $K\to \infty$, and for $N\gg 1$, the index coefficients $\mathcal{C}$ are given by
\begin{align}
  \mathcal{C}(K,P_a) \sim e^{i\pi F_\star(K,P_a)} e^{S_\text{BPS}(F_\star(K,P_a),K,P_a)}
\end{align}
Quite remarkably, this result supports the expectation that the four-parameter extremal supersymmetric Lorentzian black hole dominates the index, and thus represents a remarkable agreement between the supergravity calculation of the black hole entropy and its field theory counterpart. Conversely, if further supersymmetric solutions exist that violate the constraint (\ref{eq: non-linear rel}), as has been suggested in an analogous scenario in AdS$_5$ \cite{Markeviciute:2018yal}, then their microstates must contribute subleadingly to the index.


\section*{Acknowledgments}

I am grateful to Eric Bergshoeff, Micha Berkooz, Davide Cassani, Alejandra Castro, Robie Hennigar, Matt Heydeman, Seyed Morteza Hosseini, Seok Kim, Johannes Lahnsteiner, Gerben Oling, Domenico Orlando, Vito Pellizzani, Aaron Poole, Susanne Reffert and Ziqi Yan for helpful discussions. Special thanks go to Nick Dorey and Boan Zhao for the enjoyable collaboration \cite{Dorey:2022cfn,Dorey:2023jfw} that lead to this project. Yet more special thanks go in particular to Nikolay Bobev, Marina David, and Junho Hong, for inspiring discussions on black holes in supergravity as we completed our work \cite{Bobev:2023bxl}. Finally, thank you to Jackson Fliss and Bernardo Zan, for the inspiration behind Figure \ref{fig: ikea}.

I am supported by David Tong's Simons Investigator Award. This work has also been partially supported by the STFC consolidated grant ST/T000694/1.

\appendix

\section{Coordinate systems}
\label{app: coordinates}

The purpose of this appendix is to catalogue in excruciating detail the various sets of coordinates we use for the six-parameter black hole solution of \cite{Bobev:2023bxl}, and indeed more generally for the $(n;p_\alpha)$ orbifold solution introduced in Section \ref{subsec: orbifold}. In all, there are four:

\begin{itemize}
  \item \textbf{`Explicit solution' coordinates} $(\hat{t},\hat{r},\hat{y},\hat{z},\hat{\phi_i})$.
  
 As the name suggests, these are the coordinates that appear in \cite{Bobev:2023bxl}, in terms of which the solution is most straightforwardly expressed. The subset $(\hat{y},\hat{z},\hat{\phi}_i)$ span a topological $S^5$. We will sometimes make this more manifest by using $(\hat{\mu}_1,\hat{\mu}_2,\hat{\mu}_3)$ instead of $(\hat{y},\hat{z})$, with $\hat{\mu}_i\ge 0$ and $\hat{\mu}_1^2 + \hat{\mu}_2^2 + \hat{\mu}_3^2=1$, defined by
\begin{align}
  \hat{\mu}_i^2 = \frac{(a_i^2 - \hat{y}^2)(a_i^2 - \hat{z}^2)}{\prod_{j\neq i}(a_i^2 - a_j^2)}
\end{align}
Then, $S^5$ is realised as a $T^3$ fibration (spanned by the $\phi_i$) over a 2-simplex (spanned by the $\hat{\mu}_i$), where $T^3$ degenerates to $T^{m+1}$ on any $m$-face. The orbifold that defines the $(n;p_\alpha)$ solution is given in terms of these coordinates in (\ref{eq: orbifold action}).

\item \textbf{`Explicit asymptotics' coordinates} $(\hat{t},\hat{u},\hat{\rho}_i,\hat{\phi_i})$.

These are coordinates in which we explicitly realise the standard global coordinates on AdS$_7$ asymptotically, or more generally AdS$_7/\Z_n[p_\alpha]$ for the $(n;p_\alpha)$ solution. Note here that $\hat{\rho}_i\ge 0$ with $\hat{\rho}_1^2 + \hat{\rho}_2^2 + \hat{\rho}_3^2=1$, while the $\hat{\phi}_i\sim \hat{\phi}_i + 2\pi$ are angles.

\item \textbf{`Boosted explicit solution' coordinates} $(t,r,\xi,y,z,\phi_\alpha)$.

These coordinates are boosted\footnote{While not strictly a local Lorentz boost, in the $n\to\infty$ limit the resulting coordinates are precisely identical to those obtained by performing an infinite local Lorentz boost. The difference at finite $n$ is chosen purely to simplify the orbifold action, analogous to boosted coordinate frames in e.g. \cite{Hellerman:1997yu,Bilal:1998ys}. } relative to the hatted coordinate systems above, in such a way that the $n\to\infty$ limit corresponds to a Penrose limit in each radial slice, zooming in on a null geodesic wrapping $S^5$. Here, $\phi_\alpha \sim \phi_\alpha + 2\pi$ and $\xi\sim\xi+2\pi n$. We will sometimes replace $(y,z)$ with $(\mu_1,\mu_1,\mu_3)$, with $\mu_i\ge 0$ and $\mu_1^2 + \mu_2^2 + \hat{\mu}_3^2=1$, defined by
\begin{align}
  \mu_i^2 = \frac{(a_i^2 - y^2)(a_i^2 - z^2)}{\prod_{j\neq i}(a_i^2 - a_j^2)}
\end{align}

\item \textbf{`Boosted explicit asymptotics' coordinates} $(t,u,\xi,\rho_\alpha,\phi_\alpha)$.

Finally, we have the boosted version of the explicit asymptotics coordinates. These coordinates make manifest the Penrose limit in which the conformal boundary goes over to the pp-wave geometry. Here, $(\rho_1,\rho_2)$ are coordinates on a quarter disc of radius $n$, i.e. $\rho_\alpha \ge 0$ with $\rho_1^2 + \rho_2^2\le n$, while $\phi_\alpha \sim \phi_\alpha + 2\pi$ and $\xi \sim \xi+2\pi n$

\end{itemize}

\noindent The various transformations between these coordinate systems are summarised below in Figure \ref{fig: coordinates}.

\begin{center}
\begin{minipage}{\textwidth}
\centering
\includegraphics[width=110mm]{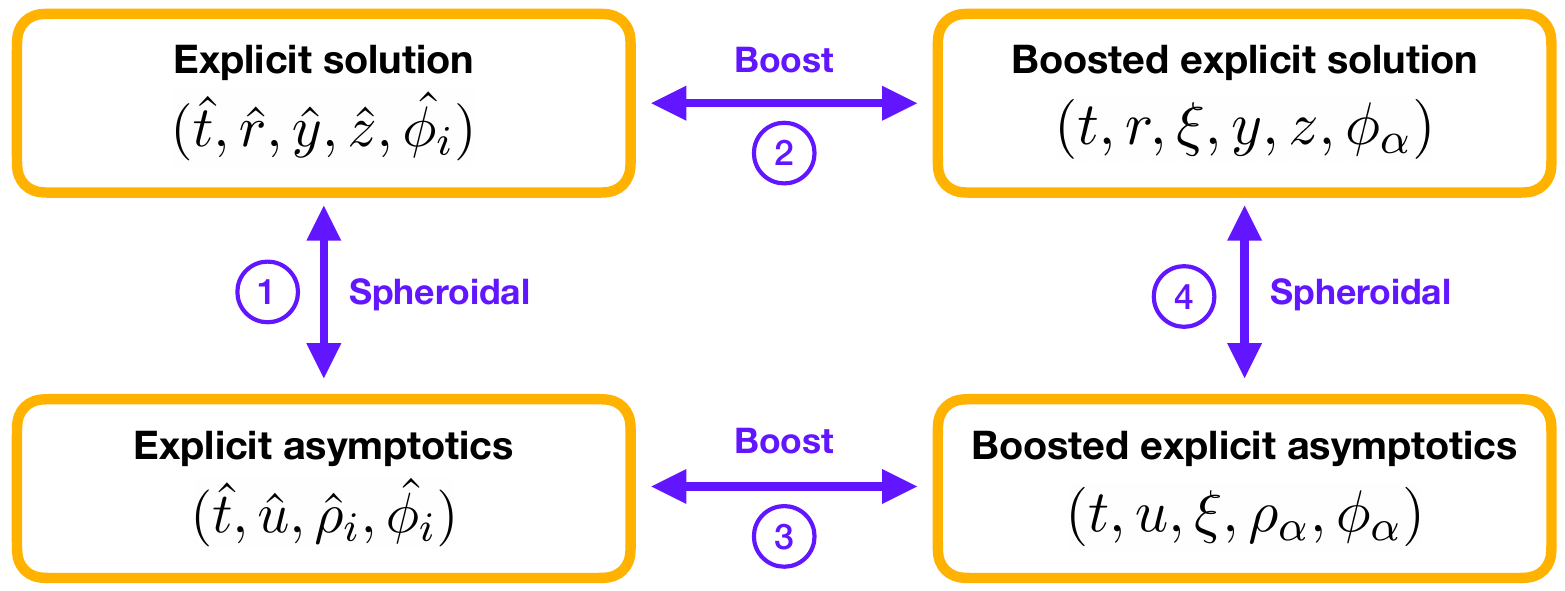}
\captionof{figure}{Different coordinate systems for the $(n;p_\alpha)$ black hole solution, and the relationships between them. The numbers refer to the explicit maps detailed below.}\label{fig: coordinates}

\end{minipage}
\end{center}
The explicit maps between each set of coordinates at finite $n$ are given by
\begin{align}
  &1.\qquad (1-g^2 a_i^2)\hat{u}^2 \hat{\rho}_i^2 = (\hat{r}^2 + a_i^2)\hat{\mu}_i^2			\nn\\
  &2. \qquad t=g\hat{t},\quad r=\hat{r},\quad \xi = n(g\hat{t} + \hat{\phi}_3),\quad y=\hat{y},\quad z=\hat{z},\quad \phi_\alpha = \hat{\phi}_\alpha		\nn\\
  &3. \qquad t = g\hat{t},\quad u = \frac{1}{\sqrt{n}}\hat{u},\quad \xi = n(g\hat{t} + \hat{\phi}_3),\quad \rho_\alpha = \sqrt{n} \hat{\rho}_\alpha,\quad \phi_\alpha = \hat{\phi}_\alpha 	\nn\\
  &4. \qquad (1-g^2 a_\alpha^2 )u^2 \rho_\alpha^2 = (r^2 + a_\alpha^2)\mu_\alpha^2,\quad n(1-g^2 a_3^2)u^2 \!\left(1-\tfrac{1}{n}(\rho_1^2 + \rho_2^2)\right) = (r^2 + a_3^2)\mu_3^2
\end{align}
where it's straightforward to check that any one of these is implied by the other three. 

The $X_7$ solution presented in Section \ref{sec: BH solution} arises in the $n\to\infty$ limit, with $a_3$ taken as in (\ref{eq: a3 behaviour}) for $\lambda$ fixed. In this limit, the ``explicit solution'' and ``explicit asymptotics'' coordinates break down, but the ``boosted explicit solution'' and ``boosted explicit asymptotics'' coordinates do not, and hence both give good coordinate systems on the $X_7$ solution. They are then related by 
\begin{align}
  (1-g^2 a_\alpha^2 )^2u^2 \rho_\alpha^2 &= \frac{g^2(a_\alpha^2 - y^2)(a_\alpha^2 - z^2)}{(a_1^2 + a_2^2-2a_\alpha^2)}(r^2 + a_\alpha^2),\nn\\
   u^2 &= \frac{\lambda(1 - g^2y^2)(1 - g^2z^2)}{g^2(1 - g^2a_1^2)(1 - g^2a_2^2)}(1+g^2 r^2)
   \label{eq: X7 coord change}
\end{align}

\section{Limiting expressions for thermodynamics}
\label{app: limits}

The aim of this Appendix is to establish formulae for the thermodynamic properties of the family of asymptotically-$X_7$ black holes presented in Section \ref{sec: BH solution} in terms of the $n\to\infty$ limit of those of the $(n;p_\alpha)$ black holes constructed in Section \ref{sec: initial solution}.

The strategy is quite simple. Once we write the $(n;p_\alpha)$ in the coordinates $(t,r,\xi,y,z,\phi_\alpha)$, it is immediately clear how to write down objects which in the limit $n\to\infty$ precisely recover the desired quantities $(E,K,J_\alpha,Q_I)$, $(\beta,\Upsilon,\Omega_\alpha, \Phi_I)$, $S$ and $I$ of the $X_7$ black hole as defined in Section \ref{subsec: thermo}. Then, using the relevant coordinate transformation set out in Appendix \ref{app: coordinates}, and keeping note of differences in convention, we construct each of these objects from the known $\hat{E}^{(n)},\dots,\hat{I}^{(n)}$, and thus determine expressions for the $X_7$ thermodynamic quantities as a limit of those of the $(n;p_\alpha)$ solution.

Let us first simply state the results. We hold fixed $(m,\lambda,a_\alpha,\delta_I)$, with $a_3$ determined as in (\ref{eq: a3 behaviour}). Then, we find
{\allowdisplaybreaks
\begin{align}
  E				&=	\lim_{n\to\infty} \bigg[g^{-1}\hat{E}^{(n)} + \hat{J}_3^{(n)}\bigg]		&&= \lim_{n\to\infty}\bigg[\frac{1}{n}\left(g^{-1}\hat{E}^{(1)} + \hat{J}_3^{(1)}\right)\bigg]		\nn\\
  K				&= \lim_{n\to\infty}\bigg[-\frac{1}{n}\hat{J}_3^{(n)}\bigg]		&&= \lim_{n\to\infty}\bigg[-\frac{1}{n^2}\hat{J}_3^{(1)}\bigg]				\nn\\
  J_\alpha		&=	 \lim_{n\to\infty}\bigg[-\hat{J}_\alpha^{(n)}\bigg]		&&= \lim_{n\to\infty}\bigg[-\frac{1}{n}\hat{J}_\alpha^{(1)}\bigg]				\nn\\
  Q_I			&=	\lim_{n\to\infty}\bigg[\frac{1}{2g}\hat{Q}_I^{(n)}\bigg]		&&= \lim_{n\to\infty}\bigg[\frac{1}{2gn}\hat{Q}_I^{(1)}\bigg]						\nn\\
  \beta			&=	\lim_{n\to\infty}\bigg[g\hat{\beta}^{(n)}\bigg]		&&= \lim_{n\to\infty}\bigg[g\hat{\beta}^{(1)}\bigg]				\nn\\
  \Upsilon		&=	\lim_{n\to\infty}\bigg[-n\left(1+g^{-1}\hat{\Omega}_3^{(n)}\right)\bigg]		&&= \lim_{n\to\infty}\bigg[-n\left(1+g^{-1}\hat{\Omega}_3^{(1)}\right)\bigg]				\nn\\
  \Omega_\alpha	&=	\lim_{n\to\infty}\bigg[-g^{-1}\hat{\Omega}_\alpha^{(n)}\bigg]		&&= \lim_{n\to\infty}\bigg[-g^{-1}\hat{\Omega}_\alpha^{(1)}\bigg]				\nn\\
  \Phi_I		&=	\lim_{n\to\infty}\bigg[2\hat{\Phi}_I^{(n)}\bigg]		&&= \lim_{n\to\infty}\bigg[2\hat{\Phi}_I^{(1)}\bigg]				\nn\\
  S				&=	\lim_{n\to\infty}\bigg[\hat{S}^{(n)}\bigg]		&&= \lim_{n\to\infty}\bigg[\frac{1}{n}\hat{S}^{(1)}\bigg]				\nn\\
  I				&=	\lim_{n\to\infty}\bigg[\hat{I}^{(n)}\bigg]		&&= \lim_{n\to\infty}\bigg[\frac{1}{n}\hat{I}^{(1)}\bigg]			
\end{align}
}%
The only one of these expressions which is not essentially immediate is the one for $E$. Recall, we define $E$ by the expression (\ref{eq: ADM energy}). After some manipulations, we find that a sufficient condition for the relation
\begin{align}
  E = \lim_{n\to\infty} \bigg[g^{-1}\hat{E}^{(n)} + \hat{J}_3^{(n)}\bigg]
  \label{eq: E limit}
\end{align}
to hold, is that in the solution of \cite{Bobev:2023bxl}, i.e. the $n=1$ solution considered here, we have
\begin{align}
  \hat{J}_3 + \frac{1}{32\pi g^3} \int_{r\to\infty,t} d^5 x \, \hat{f}(y,z) \tilde{\Omega}^{-4} \tilde{n}^c \tilde{n}^d \tilde{C}^t{}_{cbd}(\partial_{\phi_3})^b =0  
  \label{eq: J3 Komar ADM}
\end{align}
where $\hat{J}_3$ was computed in \cite{Bobev:2023bxl} by Komar integral. But (\ref{eq: J3 Komar ADM}) is precisely the statement that the Komar and ADM forms of the angular momentum $\hat{J}_3$ give the same answer. Indeed, we have verified explicitly that (\ref{eq: J3 Komar ADM}) does indeed hold, and therefore so does (\ref{eq: E limit}).

\bibliography{General7dUltra}
\bibliographystyle{JHEP}

\end{document}